\documentclass[a4paper,nobibnotes,nofootinbib]{revtex4}
\usepackage[english]{babel}
\usepackage{tabularx}
\usepackage{multirow}

\usepackage{amssymb}
\usepackage{amsmath}
\usepackage{epsfig}

\newcommand{\bc}{\begin{center}}
\newcommand{\ec}{\end{center}}

\newcommand{\reftab}[1]{Table~\ref{tab:#1}}
\newcommand{\refsec}[1]{Section~\ref{#1}}
\newcommand{\reffig}[1]{Figure~\ref{fig:#1}}
\newcommand{\refeq}[1]{(\ref{eq:#1})}

\newcommand{\SU}{\mbox{$\mathrm{SU}$}}
\newcommand{\U}{\mbox{$\mathrm{U}$}}

\newcommand{\lam}{\lambda^{\gamma}}
\newcommand{\wwgamma}{WW\gamma}
\newcommand{\wwgammagamma}{WW\gamma\gamma}
\newcommand{\zzgammagamma}{ZZ\gamma\gamma}
\newcommand{\be}{\begin{equation}}
\newcommand{\ee}{\end{equation}}
\newcommand{\bea}{\begin{eqnarray}}
\newcommand{\eea}{\end{eqnarray}}

\newcommand\lr[1]{{\left({#1}\right)}}
\newcommand{\aZerow}{a_0^W}
\newcommand{\aZeroz}{a_0^Z}
\newcommand{\aCw}{a_C^W}
\newcommand{\aCz}{a_C^Z}
\newcommand{\aZerowLambda}{a_0^W/\Lambda^2}
\newcommand{\aZerozLambda}{a_0^Z/\Lambda^2}
\newcommand{\aCwLambda}{a_C^W/\Lambda^2}
\newcommand{\aCzLambda}{a_C^Z/\Lambda^2}
\newcommand{\egamma}{E_{\gamma}}
\newcommand{\p}{\partial}
\newcommand{\units}[1]{\,\hbox{#1}}

\newcommand{\GeV}{\units{GeV}}
\def \et {E_{T}}
\newcommand{\missET}{\mbox{$\not\!\!\et$}}
\newcommand{\pt}{\mbox{$p_T$}}
\newcommand{\TeV}{\units{TeV}}
\newcommand{\fb}{\units{fb}}
\newcommand{\pb}{\units{pb}}
\newcommand{\bi}{\begin{itemize}}
\newcommand{\ei}{\end{itemize}}
\newcommand{\lumi}{\mbox{$\mathcal{L}$}}
\newcommand{\lumiunit}{\,\mbox{$\hbox{cm}^{-2}\hbox{s}^{-1}$}}
\newcommand{\twosidep}[1]{\stackrel{\leftrightarrow}{\p^{#1}}}
\newcommand{\pom}{\mathbb{P}}
\newcommand{\reg}{\mathbb{R}}

\newcommand{\dkap}{\Delta\kappa^{\gamma}}
\newcommand{\amp}{{\mathcal A}}

\renewcommand{\d}{\mathrm{d}}

\newcommand{\rad}{\units{rad}}
\newcommand{\invpb}{\mbox{$\units{pb}^{-1}$}}
\newcommand{\invfb}{\mbox{$\units{fb}^{-1}$}}
\newlength{\threepicwidth}
\newlength{\twopicwidth}
\newlength{\picwidthbig}
\newlength{\picwidthsq}
\newlength{\picwidth}
\newlength{\smallpicwidth}
\begin{document}
\title{Anomalous quartic WW$\gamma\gamma$, ZZ$\gamma\gamma$,
and trilinear WW$\gamma$ couplings in two-photon processes 
at high luminosity at the LHC}
\author{E. Chapon}\email{emilien.chapon@cea.fr}
\affiliation{CEA/IRFU/Service de physique des particules, CEA/Saclay, 91191 
Gif-sur-Yvette cedex, France}
\author{O. Kepka}\email{kepkao@fzu.cz}
\affiliation{CEA/IRFU/Service de physique des particules, CEA/Saclay, 91191
Gif-sur-Yvette cedex, France}
\affiliation{IPNP, Faculty of Mathematics and Physics,
Charles University, Prague} 
\affiliation{Center for Particle Physics, Institute of Physics, Academy of Science, Prague} 
\author{C. Royon}\email{royon@hep.saclay.cea.fr}
\affiliation{CEA/IRFU/Service de physique des particules, CEA/Saclay, 91191 
Gif-sur-Yvette cedex, France}

%%%%%%%%%%%%%%%%%%%%%%%%%%%%%%%%%%%%%%%%%%%%%%%%%%%%%%%%%%%%%%%%%%
%%%%%%%%%%%%%%%%%%%%%%%%   Abstract   %%%%%%%%%%%%%%%%%%%%%%%%%%%%
%%%%%%%%%%%%%%%%%%%%%%%%%%%%%%%%%%%%%%%%%%%%%%%%%%%%%%%%%%%%%%%%%%
\begin{abstract}
We study the $W/Z$ pair production via two-photon exchange at the LHC and
give the sensitivities on trilinear and quartic gauge anomalous couplings 
between photons and $W/Z$ bosons for an integrated luminosity of 30 and 
200 fb$^{-1}$. For simplicity and to obtain lower backgrounds, only the leptonic
decays of the electroweak bosons are considered.
\end{abstract}
\maketitle

In the Standard Model (SM) of particle physics, the couplings of fermions and 
gauge bosons are constrained by the gauge symmetries of the Lagrangian.
The measurement of $W$ and $Z$ boson pair productions via the exchange of
two photons  
allows to provide directly stringent tests
of one of the most important and least understood
mechanism in particle physics, namely the
electroweak symmetry breaking~\cite{stirling}. The non-abelian gauge nature of the SM
predicts the existence of quartic couplings
$WW\gamma \gamma$
between the $W$ bosons and the photons which can be probed directly at the 
Large Hadron Collider (LHC) at CERN.
The quartic coupling $ZZ\gamma \gamma$ is not present in the
SM. 

The quartic couplings test 
more generally new physics which couples to electroweak bosons.
Exchange of heavy particles beyond the SM might manifest itself as a
modification of the quartic couplings appearing in contact 
interactions~\cite{higgsless}. It is
also worth noticing that in the limit of infinite Higgs masses, or in Higgs-less
models~\cite{higgsless}, new structures not present in the tree level Lagrangian
appear in the quartic $W$ coupling. For
example, if the electroweak breaking mechanism does not manifest itself in the
discovery of the Higgs boson at the LHC or supersymmetry, the presence of
anomalous couplings might be the first evidence of new
physics in the electroweak sector of the SM.

Two-photon physics is thus a significant enhancement of the LHC physics 
program~\cite{piotr}. It allows to study the Standard Model 
in a unique way at an hadron collider through exchange of photons. This 
paper focuses on two applications of the diboson production in two-photon 
events. First we propose a measurement of the $pp\rightarrow pWW\!p$ cross 
section with the use of forward detectors to tag the intact protons, that leave
the interaction intact at small angles. 
Second, we explore the sensitivities to anomalous quartic $\wwgammagamma$, 
$\zzgammagamma$ (QGC) and triple $\wwgamma$ (TGC) gauge couplings. 
Benefiting from the enhancement of the cross section when anomalous couplings are 
considered, the study
of QGC sensitivities is performed for two values of integrated luminosity,
namely 30 and 200 fb$^{-1}$
at the LHC at the nominal center-of-mass energy of 14 TeV. To
simplify the study and reduce the amount of background, we restrict ourselves 
to consider only the leptonic decays of the $W$ and $Z$ bosons.

The plan of this paper is as follows. The first section is
dedicated to the theoretical framework of the photon induced
processes. The second section describes the effective Lagrangians of the anomalous
 triple and quartic couplings which we are intending to study. 
In the third section, we discuss the implementation of the two-photon and diffractive
processes inside the Forward Physics Monte Carlo (FPMC) which we used
to generate all our signal and background. In section four, we describe 
the methods to extract the diffractive and two-photon events with forward detectors
at the LHC.  The possibility to observe SM $W$-pair production
via two-photon exchange is discussed in the fifth  section and the 
section six is dedicated to the derivation of the sensitivity to $\gamma
\gamma WW$ or $\gamma \gamma ZZ$ anomalous quartic
couplings at the LHC. In the last
section, we discuss the sensitivity to $\gamma WW$ triple gauge anomalous
couplings.

%%%%%%%%%%%%%%%%%%%%%%%%%%%%%%%%%%%%%%%%%%%%%%%%%%%%%%%%%%%%%%%%%
\section{Two-photon Exchange in the Standard Model}
%%%%%%%%%%%%%%%%%%%%%%%%%%%%%%%%%%%%%%%%%%%%%%%%%%%%%%%%%%%%%%%%%%
In this section, we first describe the theoretical framework of photon induced processes before 
focusing on the $W$-pair production through two-photon exchange which we intend to study.

%%%%%%%%%%%%%%%%%%%%%%%%%%%%%%%%%%%%%%%%%%%%%%%%%%%%%%%%%%%%%%%%%%
\subsection{Two-photon production cross section}
%%%%%%%%%%%%%%%%%%%%%%%%%%%%%%%%%%%%%%%%%%%%%%%%%%%%%%%%%%%%%%%%%%

\begin{figure}
\includegraphics[width=\twopicwidth]{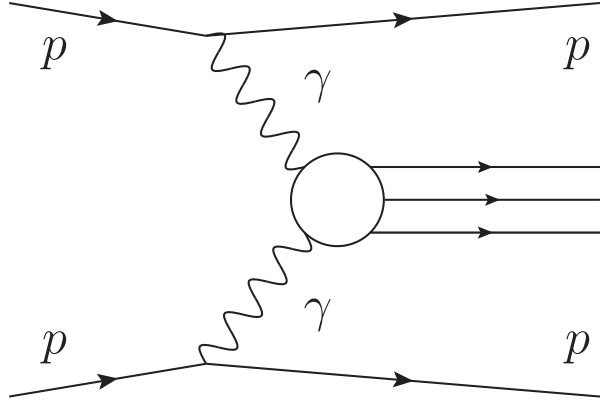}	
\caption{Sketch diagram showing the two-photon production of a central system. Unaltered protons leave
the interaction at very small angles $\lesssim100\,\mu$rad and the central system is produced alone in the central detector without any proton remnants.}
\label{fig:photon_photon.eps}
\end{figure}

Two-photon production in $pp$ collision is described in the framework of the Equivalent Photon Approximation 
(EPA)~\cite{Budnev}. The almost real photons (low photon virtuality $Q^2=-q^2$) are 
emitted by the incoming protons producing an object $X$, $pp\rightarrow pXp$, 
through two-photon exchange $\gamma\gamma\rightarrow X$, see Figure~\ref{fig:photon_photon.eps}.
The photon spectrum of virtuality $Q^2$ and
energy $E_{\gamma}$ is proportional to the Sommerfeld fine-structure constant $\alpha$  and reads
\begin{equation}
\d N = \frac{\alpha}{\pi}\frac{\d E_{\gamma}}{E_{\gamma}}\frac{\d Q^2}{Q^2}
 	 \left[ \left(1-\frac{E_{\gamma}}{E}\right)\left(1-\frac{Q^2_{min}}{Q^2}\right)F_E +
	         \frac{E_{\gamma}^2}{2E^2}F_M\right]
\label{eq:flux}
\end{equation}
where $E$ is the energy of the incoming proton of mass $m_p$, $Q^2_{min}\equiv
m^2_p E^2_{\gamma}/[E(E-E_{\gamma})]$ the photon minimum virtuality allowed by
kinematics and $F_E$ and $F_M$ are functions of the electric and magnetic form factors. They read
in the dipole approximation~\cite{Budnev}
\begin{equation}
F_M=G^2_M  \qquad F_E=(4m_p^2G^2_E+Q^2G^2_M)/(4m_p^2+Q^2)\qquad G^2_E=G^2_M/\mu_p^2=(1+Q^2/Q^2_0)^{-4}
\label{eq:newera:elmagform}
\end{equation}
The magnetic moment of the proton is $\mu_p^2=7.78$ and the fitted scale $Q^2_0=0.71\,\GeV^2$.
Electromagnetic form factors are steeply falling  as a function of $Q^2$. That is the 
reason why the two-photon cross section can be factorized into the sub-matrix element and  two photon 
fluxes. To obtain the production cross section, the photon fluxes are first integrated over $Q^2$
\be
f(E_\gamma)=\int^{Q^2_{max}}_{Q^2_{min}}\frac{\d N}{\d E_\gamma \d Q^2} \d Q^2
\label{sm:flux_q2}
\ee
up to a sufficiently 
large value of $Q^2_{max}\thickapprox2-4\GeV^2$. The result can be written as 

\begin{equation}
\d N(\egamma)=\frac{\alpha}{\pi}\frac{\d\egamma}{\egamma}
\lr{1-\frac{\egamma}{E}}
\left[\varphi\lr{\frac{Q^2_{max}}{Q^2_0} }-\varphi\lr{\frac{Q^2_{min}}{Q^2_0} }\right]
\label{app:eq:budnev}
\end{equation}
where the function $\varphi$ is defined as 

\begin{eqnarray}
\varphi(x)&=&(1+ay)
\left[%
-\ln(1+x^{-1})+\sum_{k=1}^{3}\frac{1}{k(1+x)^{k}}
\right]
+ \frac{(1-b)y}{4x(1+x)^3}\nonumber\\
&+&c(1+\frac{y}{4})
\left[%
\ln\frac{1+x-b}{1+x}+\sum_{k=1}^{3}\frac{b^k}{k(1+x)^{k}}
\right]
\label{phi}
\end{eqnarray}
where
\begin{eqnarray}
 y&=&\frac{\egamma^2}{E(E-\egamma)} \nonumber \\
 a&=&\frac{1}{4}(1+\mu_p^2)+\frac{4m_p^2}{Q^2_0}\approx 7.16\nonumber\\
 b&=&1-\frac{4m_p^2}{Q_0^2}\approx -3.96\nonumber \\
 c&=&\frac{\mu_p^2-1}{b^4}\approx 0.028
\end{eqnarray}

Note that the formula for the $Q^2$-integrated 
photon flux was quoted incorrectly several  times in the literature. There is a 
sign error in the original paper in Ref.~\cite{Budnev} in the second term
of $\varphi(x)$ in Equation~\ref{phi}.  
Moreover, in \cite{Boonekamp:2007iu} there is another typesetting error leading to wrong second and last terms.
\begin{figure}
\centering
\includegraphics[width=1\twopicwidth]{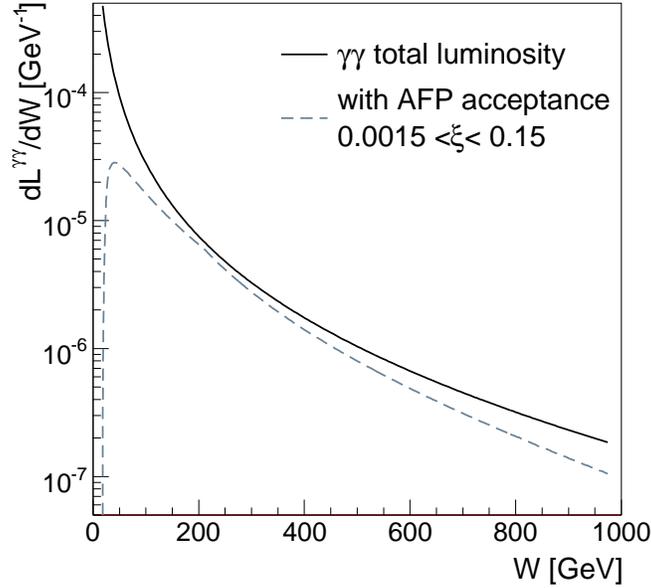}
\caption{Relative effective $\gamma\gamma$ luminosity in $pp$ collisions at $14\units{TeV}$
as a function of the  two-photon invariant mass. The maximal virtualities of the
emitted photons are set to $Q^2_{max}=2\,\GeV^2$. The dashed curve
shows the photon spectrum within the ATLAS or CMS forward detector acceptance (discussed in section \ref{sec:selection}).}
\label{fig:luminosity}
\end{figure}

The contribution to the integral above $Q^2_{max}\thickapprox2\GeV^2$  is very small. 
The $Q^2$-integrated photon flux also falls rapidly as a function of the photon energy $E_{\gamma}$ 
which implies that the two-photon production is dominant at small masses $W\approx2\sqrt{E_{\gamma1}E_{\gamma2}}$. 
Integrating the product of the
photon fluxes $f(E_{\gamma1})\cdot f(E_{\gamma2})\cdot \d E_{\gamma1}\cdot 
\d E_{\gamma2}$ from both protons over the photon  energies while keeping
the two-photon invariant mass fixed to $W$, one obtains the two-photon effective
luminosity spectrum $\d L^{\gamma\gamma}/\d W$.

The effective $\gamma \gamma$ luminosity 
is shown in \reffig{luminosity} as a function of the mass $W$ in full line. 
The production of heavy objects is particularly interesting
at the LHC where new particles could be produced in a very clean environment. 
The production rate of massive objects is however limited by the photon 
luminosity at high invariant mass. The integrated two-photon 
luminosity above $W>W_0$ for $W_0=23\GeV,\ 2\times m_W\thickapprox160\GeV$, 
and $1\TeV$ is respectively $1\%$, $0.15\%$ and $0.007\%$ of the luminosity 
integrated over the whole mass spectrum. The luminosity spectrum was
calculated using the upper virtuality bound $Q^2_{max}=2\,\GeV^2$ using 
numerical integration. The luminosity spectrum 
within the proposed forward detector acceptance to detect the intact
protons $0.0015<\xi<0.15$ is also shown in the figure (it is calculated in the limit of low
$Q^2$, thus setting $E_{\gamma}=\xi E$). 

Using the effective relative photon
luminosity $\d L^{\gamma\gamma} \slash \d W$, the total cross section reads 
\begin{equation}
      \frac{\d\sigma}{\d\Omega}=\int\frac{\d \sigma_{\gamma\gamma\rightarrow X}
      (W)}{\d\Omega}\frac{\d L^{\gamma\gamma}}{\d W}\d W 
\label{eq:sm:totcross}
\end{equation}
where $\d \sigma_{\gamma\gamma\rightarrow X}/\d\Omega$ denotes the differential 
cross section of the sub-process $\gamma\gamma\rightarrow X$, dependent on the invariant mass of the two-photon system.

%%%%%%%%%%%%%%%%%%%%%%%%%%%%%%%%%%%%%%%%%%%%%%%%%%%%%%%%%%%%%%%%%%
\subsection{$W$ pair production via photon exchanges}
%%%%%%%%%%%%%%%%%%%%%%%%%%%%%%%%%%%%%%%%%%%%%%%%%%%%%%%%%%%%%%%%%%

\begin{figure}
\parbox{\textwidth}{
\parbox{\threepicwidth}{\includegraphics[width=\threepicwidth]{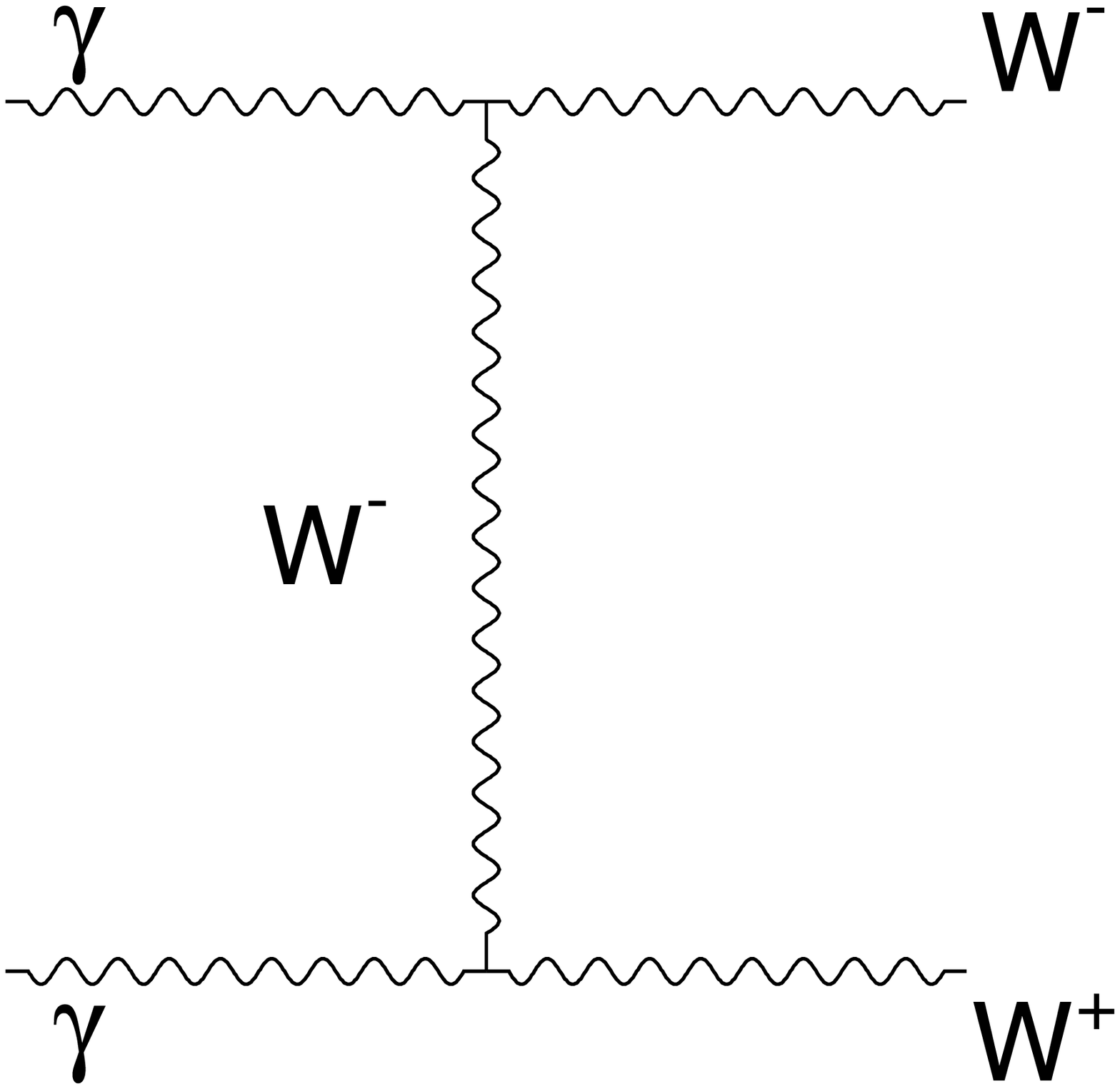}\\ \centering (a)}
\parbox{\threepicwidth}{\includegraphics[width=\threepicwidth]{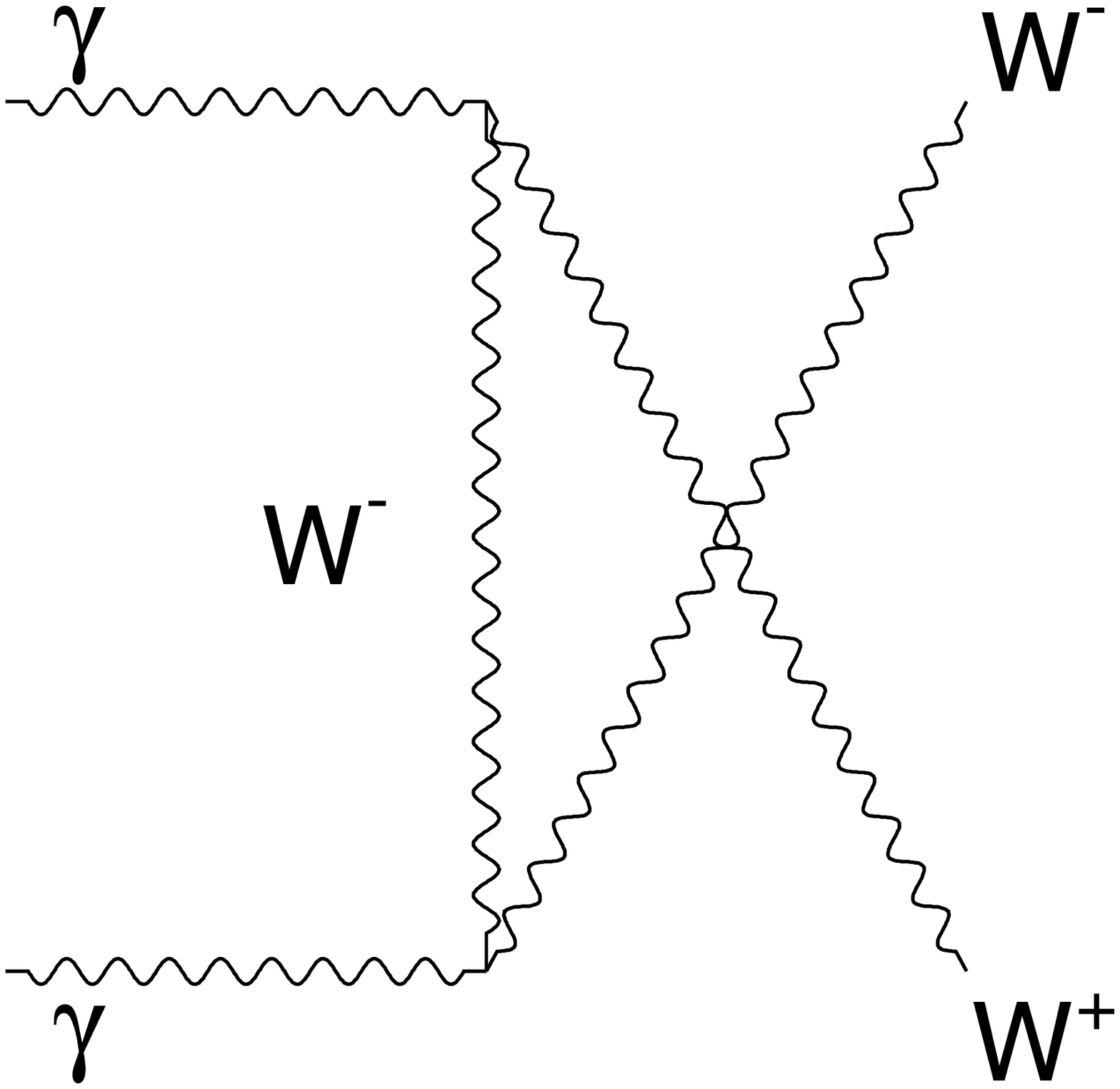}\\ \centering (b)}
\parbox{\threepicwidth}{\includegraphics[width=\threepicwidth]{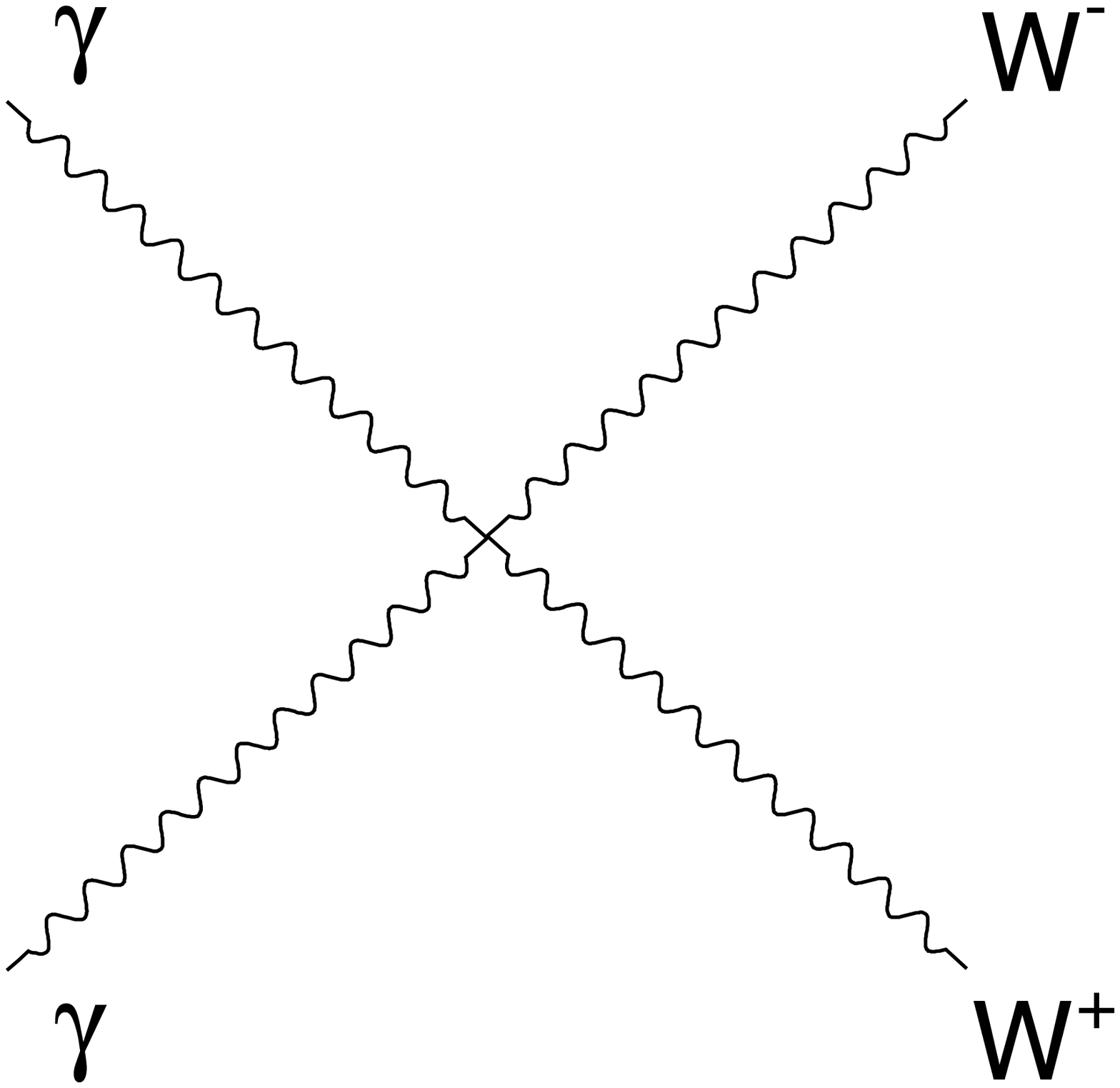}\\ \centering (c)}
}
\caption{Feynman diagrams of SM processes that contribute to the 
$\gamma\gamma\rightarrow WW$ scattering amplitude in the lowest order 
perturbation series with a coupling $e^2$. The trilinear couplings of 
strength $e$ are involved in diagrams a) and b) and the direct quartic 
coupling of strength $e^2$ in diagram c).}
\label{fig:anom:diagramsSM}
\end{figure}

The process that we intend to study is the $W$ pair production induced by the 
exchange of two photons as shown in \reffig{anom:diagramsSM}. It is a pure QED process
in which the decay products of the $W$ bosons are measured in the central 
detector and the scattered protons leave intact in
the beam pipe at very small angles, contrary to inelastic collisions. Since 
there is no proton remnant the process is purely exclusive; only $W$ decay products 
populate the central detector, and the intact protons can be detected in
dedicated detectors located along the beam line far away from the interaction
point.

Considering the 
interactions with at least one photon, three-boson $\wwgamma$, and 
four-boson  $\wwgammagamma$ interactions read
\begin{eqnarray}
\lumi_{WW\gamma} & = & -ie(A_\mu W^-_\nu \twosidep{\mu} W^{+\nu}+W_\mu^-W^+_\nu\twosidep{^\mu} A^\nu
+W^+_\mu A_\nu\twosidep{\mu}W^{-\nu})\label{eq:anom:lagrww1}\\
\lumi_{WW\gamma\gamma} & = & -e^2(W^{-}_\mu W^{+\mu}A_\nu A^\nu-W_\mu^-A^\mu W^+_\nu A^\nu)\label{eq:anom:lagrww2}
\end{eqnarray}
where the asymmetric derivative has the form 
$X\twosidep{^\mu}Y=X\p^{\mu}Y-Y\p^{\mu}X$.

The production of $Z$ bosons via two-photon exchange is forbidden in the lowest 
order perturbation theory because
neither the $Z$ boson nor the photon carries an electric or weak charge. On 
the other hand, the $W$ boson can be produced
in pairs. In this case, both the triple gauge $\wwgamma$ (with $t-$ and 
$u-$channel exchange) and the quartic gauge $\wwgammagamma$ boson interactions 
must be included as shown in \reffig{anom:diagramsSM}.

\par

%%%%%%%%%%%%%%%%%%%%%%%%%%%%%%%%%%%%%%%%%%%%%%%%%%%%%%%%%%%%%%%%%%%
%\subsection{Tree level unitarity and divergence cancellation }
%%%%%%%%%%%%%%%%%%%%%%%%%%%%%%%%%%%%%%%%%%%%%%%%%%%%%%%%%%%%%%%%%%

In the $\gamma\gamma\rightarrow WW$ process, the fundamental property of 
divergence cancellations in the SM at high
energy is directly effective. A necessary condition for the renormalizibility
of the Standard Model at all orders is the so called ``tree unitarity" demanding 
that the unitarity is only minimally (logarithmically) violated in any fixed order 
of the perturbation series~\cite{TreelevUnit1,TreelevUnit2}.
For the binary process of  $W$ pair production in particular, the tree level unitarity implies that the 
scattering amplitude $\gamma\gamma\rightarrow WW$ should be a constant or vanish in the high energy
limit. In the SM, this condition is indeed satisfied due to the cancellation between $t$-, $u$-channel and direct quartic diagrams.
\par 
The cross section is constant in the high energy limit.  The leading 
order differential formula for the $\gamma\gamma\rightarrow WW$ process
is a function of the Mandelstam variables $s,t,u$ and the mass of the vector 
boson $W$
\cite{electroweakCorrections}

\begin{equation}
\frac{\d\sigma}{\d\Omega}=\frac{3\alpha^2\beta}{2s}\left\{1
-\frac{2s(2s+3M_W^2)}{3(M_W^2-t)(M^2_W-u)}
+\frac{2s^2(s^2+3M_W^4)}{3(M_W^2-t)^2(M_W^2-u)^2}
\right\}
\label{eq:anom:wwprod}
\end{equation}
where $\beta=\sqrt{1-4M_W^2/s}$ is the velocity of the $W$ bosons. For 
$s\rightarrow \infty$ the total cross section 
is $\sigma_{\mathrm{tot}}=80.8\pb$.

\par
Measuring the $\gamma\gamma\rightarrow WW$ scattering process at the LHC is 
therefore interesting not only because we can use the hadron-hadron machine as 
the photon-photon collider with a clean collision environment without beam 
remnants, but also because it provides a very clear test of the Standard Model 
consistency in a rather textbook process. 

\par
The cross section of the $pp\rightarrow p WW p$ process which proceeds through 
two-photon exchange is effectively calculated as a convolution \refeq{sm:totcross} of the 
two-photon luminosity and the total cross section $\gamma\gamma\rightarrow WW$ \refeq{anom:wwprod}.
The total two-photon cross section is 95.6\fb. 
\par  Since the virtuality of the photon is very 
close to zero, the electromagnetic coupling appearing in the interaction 
Lagrangians in Equations~\refeq{anom:lagrww1} and \refeq{anom:lagrww2} is evaluated at the 
scale $Q^2=0$; the electromagnetic fine-structure constant therefore takes the 
value $\alpha=1/137$. Note that the above mentioned total cross section is 
different from the usually presented value of 
108\fb{} (see \cite{Pierzchala:2008xc} for example) by about 10\%.
This is due to the fact that the authors considered the fixed value of the electromagnetic 
coupling of 1/129 at the scale of the $W$ mass.
In fact, the photon virtuality should be taken as the scale and not the mass 
of the $W$. In the Landau gauge, 
the invariant charge is driven by the self-energy insertion into the photon 
propagator only (and not by the vertex correction)~\cite{misha}. In the 
propagator we have to take the photon virtuality as the scale, which is very 
small. The total two-photon cross section is therefore $\sigma_{}=95.6\fb$. This value 
has to be corrected for the survival probability factor 0.9.

%%%%%%%%%%%%%%%%%%%%%%%%%%%%%%%%%%%%%%%%%%%%%%%%%%%%%%%%%%%%%%%%%%
\section{$W$ and $Z$ photon quartic and trilinear anomalous couplings}
%%%%%%%%%%%%%%%%%%%%%%%%%%%%%%%%%%%%%%%%%%%%%%%%%%%%%%%%%%%%%%%%%%
 The two-photon production of dibosons 
is very suitable to test the electroweak theory because it allows to probe 
trilinear and quartic boson couplings. The test is based on deriving the 
sensitivities with a counting experiment to parameters (coupling strengths) of new auxiliary interaction 
Lagrangians  
added to the SM, to simulate low energetic effects of some Beyond Standard Model (BSM) theories whose 
typical scales (e.g the typical new particle masses) are beyond the reach of 
the LHC energies.
In this section, we give the theoretical implementation of quartic
and trilinear anomalous
couplings between the $W$ or $Z$ boson and the photon in the FPMC generator.

%%%%%%%%%%%%%%%%%%%%%%%%%%%%%%%%%%%%%%%%%%%%%%%%%%%%%%%%%%%%%%%%%%
\subsection{Effective quartic anomalous Lagrangian}
%%%%%%%%%%%%%%%%%%%%%%%%%%%%%%%%%%%%%%%%%%%%%%%%%%%%%%%%%%%%%%%%%%

%%%%%%%%%%%%%%%%%%%%%%%%%%%%%%%%%%%%%%%%%%%%%%%%%%%%%%%%%%%%%%%%%%
\subsubsection{Construction of new quartic anomalous operators}
%%%%%%%%%%%%%%%%%%%%%%%%%%%%%%%%%%%%%%%%%%%%%%%%%%%%%%%%%%%%%%%%%%
The boson self-interaction  in the SM is completely 
derived from the underlying
$\SU(2)_L\times \U_Y(1)$ local symmetry. New vector boson fields are added to 
the Lagrangian to guarantee the invariance under this symmetry and their 
self-interactions emerge from the vector boson kinetic terms.

The vector boson masses are, however, more deeply linked with the Higgs field 
and the vacuum symmetries. The symmetry O(4) of the Higgs potential 
$V(\Phi)= -\mu^2\Phi^{\dagger}\Phi+\lambda(\Phi^\dagger\Phi)^2$ 
is in fact larger than the 
required 
$\SU(2)\times \U(1)$. It is known that the symmetry O(4) is locally isomorphic 
to $\mathrm{O}(4)\simeq \SU(2)\times\SU(2)$. When the symmetry is 
spontaneously broken and 
one particular vacuum $\Phi_U$ is chosen, the vacuum symmetry is reduced. The 
vacuum is invariant under $\SU(2)$ only. The weak isospin generators 
$\vec{\tau}/2$ corresponding to the broken symmetry
constitute a triplet with respect to the vacuum symmetry sub-group. Very 
interestingly, this vacuum symmetry controls the value of the 
$\rho$ parameter
\be
\rho=\frac{M^2_W}{M^2_Z\cos^2\theta_W}
\ee
and is usually called the custodial $\SU(2)_C$ symmetry. 
The SM value of the parameter is  $\rho=1$ and it was very well confirmed 
experimentally  (taking $m_W=80.396\pm0.025$, $m_Z=91.1876\pm0.021$, and 
$\sin^2\theta_W=0.231\mp0.00023$ as in \cite{Amsler:2008zzb}, we obtain 
$\rho=1.011\pm0.001$ so it is known with a precision better than 1\%). 
In models with higher Higgs multiplets, $\rho$ can significantly differ from 1. 
We will assume that this symmetry holds also in more general theories which we 
are about to parameterize and construct new effective Lagrangian terms in such 
a way to obey the deeper $\SU(2)_C$ symmetry which is tightly linked with the 
precisely measured value of the $\rho$ parameter.

The boson self-interactions in the SM (including their kinetic terms) can be 
conveniently represented by $-\frac{1}{4}
 W_{\mu\nu}\cdot W^{\mu\nu}$ where the vector
\be
\vec{W}_\alpha=\left (
 \begin{array}{c}
\frac{1}{\sqrt{2}}(W^+_\alpha + W^-_\alpha) \\
\frac{i}{\sqrt{2}}(W^+_\alpha - W^-_\alpha) \\
Z_\alpha/\cos\theta_W
\label{eq:anom:triplet}
\end{array}
\right)
\ee
is a triplet of the custodial $\SU(2)_C$ symmetry. The field tensor for $W$ 
bosons appearing in the product is 
$\vec{W}_{\mu\nu}=\p_\mu \vec{W}_\nu - \p_\nu \vec{W}_\mu  + 
g \vec{W}_\mu\times\vec{W}_\nu$.

In the following, the parameterization of the quartic couplings
based on \cite{Belanger:1992qh} is adopted. We concentrate on the lowest order 
dimension operators which have
the correct Lorentz invariant structure and obey the $\SU(2)_C$ 
custodial symmetry in order to fulfill the stringent experimental bound on the 
$\rho$ parameter. Also, the $\U(1)_Q$ gauge symmetry for those operators 
which involve photons, is required.

There are only two four-dimension operators: 
\begin{eqnarray}
\lumi^0_4&=&\frac{1}{4}g_0 g_W(\vec{W}_\mu\cdot\vec{W}^\mu)^2\nonumber\\
\lumi^C_4&=&\frac{1}{4}g_C g_W(\vec{W}_\mu\cdot\vec{W}_\nu)(\vec{W}^\mu\cdot\vec{W}^\nu)
\end{eqnarray}
They are parameterized by the corresponding couplings $g_0$ and $g_C$.
Using the explicit form of the $\SU(2)_{C}$ triplet we see 
that these Lagrangians do not involve
photons. Clearly, it is not possible to construct any operator of dimension 5 
since an even number of Lorentz indices is needed to contract the field 
indices. Thus the lowest order interaction
Lagrangians which involve two photons are dim-6 operators. There are two of 
them:
\begin{eqnarray}
\lumi^{0}&=&-\frac{\pi\alpha}{4\Lambda^2}a_0 F_{\alpha\beta}F^{\alpha\beta}(\vec{W}_\mu\cdot\vec{W}^\mu)\\
\lumi^{C}&=&-\frac{\pi\alpha}{4\Lambda^2}a_C F_{\alpha\mu}F^{\alpha\nu}(\vec{W}^\mu\cdot\vec{W}^\nu)
\end{eqnarray}
parameterized with new coupling constants $a_0$, $a_C$, and the fine-structure 
constant $\alpha=e^2/(4\pi)$. The new scale $\Lambda$ is introduced
so that the Lagrangian density has the correct dimension four and is 
interpreted as the typical mass scale of  new
physics. Expanding the above formula using the definition of the $\SU(2)_C$ 
triplet \label{eq:anom:triple} and expressing the product 
\be
\vec{W}_\mu\cdot\vec{W}_\nu = 2\left(W^+_\mu W^-_\nu+
\frac{1}{2\cos^2\theta_W}Z_\mu Z_\nu\right)
\ee
we arrive at the following expression for the effective quartic Lagrangian
 \begin{eqnarray}
     \mathcal{L}_6^0 &=& \frac{-e^2}{8} \frac{\aZerow}{\Lambda^2} F_{\mu\nu} 
     F^{\mu\nu} W^{+\alpha} W^-_\alpha 
     - \frac{e^2}{16\cos^2 \theta_W} \frac{a^Z_0}{\Lambda^2} F_{\mu\nu} 
     F^{\mu\nu} Z^\alpha
Z_\alpha\nonumber \\
     \mathcal{L}_6^C & = & \frac{-e^2}{16} \frac{\aCw}{\Lambda^2} 
     F_{\mu\alpha} F^{\mu\beta} (W^{+\alpha} W^-_\beta + W^{-\alpha} 
     W^+_\beta) 
	- \frac{e^2}{16\cos^2 \theta_W} \frac{a^Z_C}{\Lambda^2} 
	F_{\mu\alpha} F^{\mu\beta} Z^\alpha Z_\beta
\label{eq:anom:lagrqgc}
\end{eqnarray}
In the above formula, we allowed the $W$ and $Z$ parts of the Lagrangian to 
have specific couplings, i.e.
 $a_0\rightarrow (\aZerow$, $a^Z_0$) and similarly $a_C\rightarrow(\aCw$, 
 $\aCz$). From the structure of 
$\mathcal{L}_6^0$ in which the indices of photons and $W$ are decoupled, we 
see that this Lagrangian can be interpreted as the exchange of a neutral 
scalar particle whose propagator does not have any Lorentz index.
A such Lagrangian density conserves $C-$, $P-$, and $T-$parities separately 
and hence represents the most natural
extension of the SM.

The current best experimental 95\% CL limits on the above anomalous 
parameters come from the OPAL Collaboration where the quartic couplings were
measured in $e^+e^-\rightarrow W^+W^-\gamma$, 
$e^+e^-\rightarrow \nu\bar{\nu}\gamma\gamma$ (for $WW\gamma\gamma$ anomalous
couplings), and $e^+e^-\rightarrow q\bar{q} \gamma\gamma$ (for 
$ZZ\gamma\gamma$ couplings) at center-of-mass energies up to 209\GeV{}. The 
corresponding $95\%$ confidence level limits on the anomalous coupling 
parameters were found \cite{LEPlimitsQGC}
\begin{eqnarray}
-0.020 \GeV^{-2} < &\aZerowLambda& < 0.020\GeV^{-2} \nonumber \\
-0.052 \GeV^{-2} < &\aCwLambda& <0.037\GeV^{-2}\nonumber\\
-0.007 \GeV^{-2} < &\aZerozLambda& <  0.023 \GeV^{-2}\nonumber\\
-0.029 \GeV^{-2} < &\aCzLambda& < 0.029\GeV^{-2} 
\label{eq:anom:qgclimits}
\end{eqnarray}
On the other hand, there has not been any direct constraint on the anomalous 
quartic couplings reported from the Tevatron so far.

\begin{figure}
\centering 
\includegraphics[width=\twopicwidth]{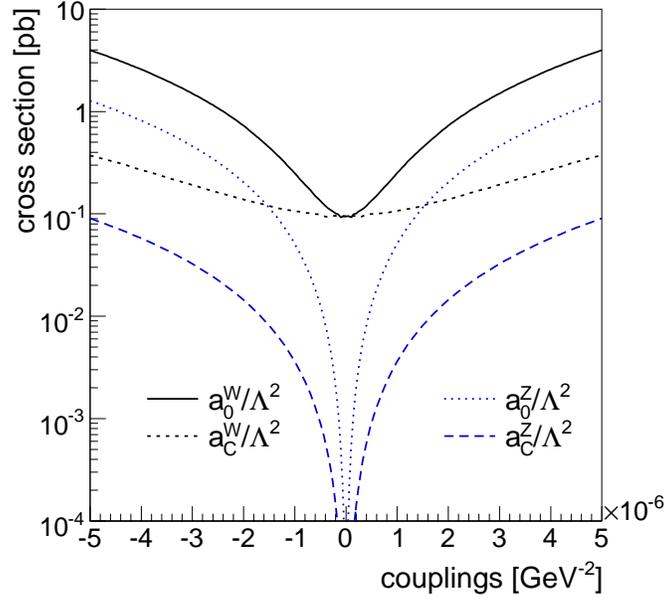}
\caption{Enhancement of the $pp\rightarrow pWWp$ and $pp\rightarrow pZZp$ cross section at $\sqrt{s}=14\TeV$ with quartic-boson anomalous couplings $\aZerow$, $\aCw$, and $\aZeroz$, $\aCz$  from the SM values 95.6\fb{} and 0, respectively. The survival probability factor is not included.}
\label{fig:anom:totalxsectionQGC}
\end{figure}

\begin{figure}
\centering
\includegraphics[height=\twopicwidth]{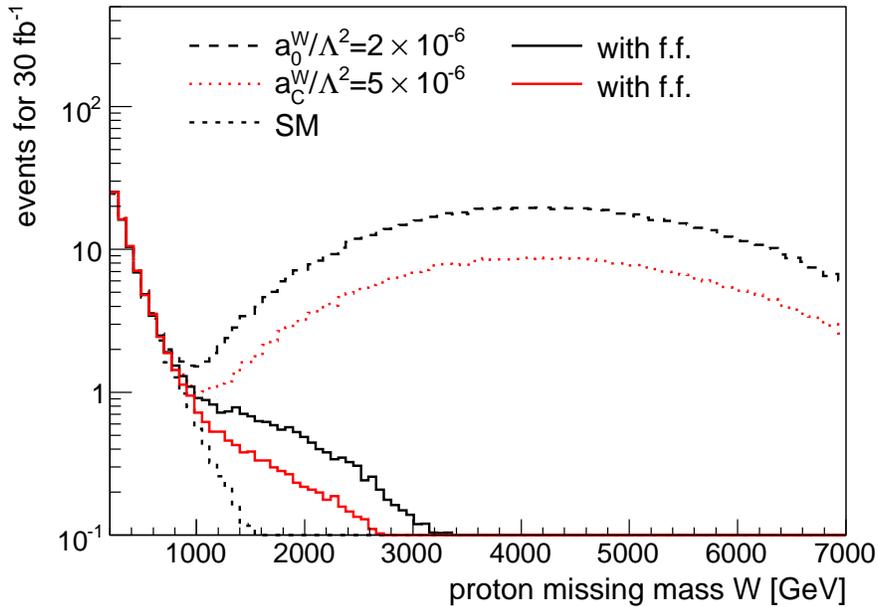}
\caption{Missing mass distribution showing the effect of the form factor 
\refeq{anom:formfactor} on the cross section. The signal due to the anomalous 
coupling appears for masses $W>800\GeV$. Both leptons are in the detector 
acceptance and above 
$\pt>10\GeV$.}
\label{fig:anom:w2form}
\end{figure}

%%%%%%%%%%%%%%%%%%%%%%%%%%%%%%%%%%%%%%%%%%%%%%%%%%%%%%%%%%%%%%%%%%
\subsubsection{Coupling form factors}
%%%%%%%%%%%%%%%%%%%%%%%%%%%%%%%%%%%%%%%%%%%%%%%%%%%%%%%%%%%%%%%%%%

\begin{figure}
\centering
\includegraphics[height=\twopicwidth]{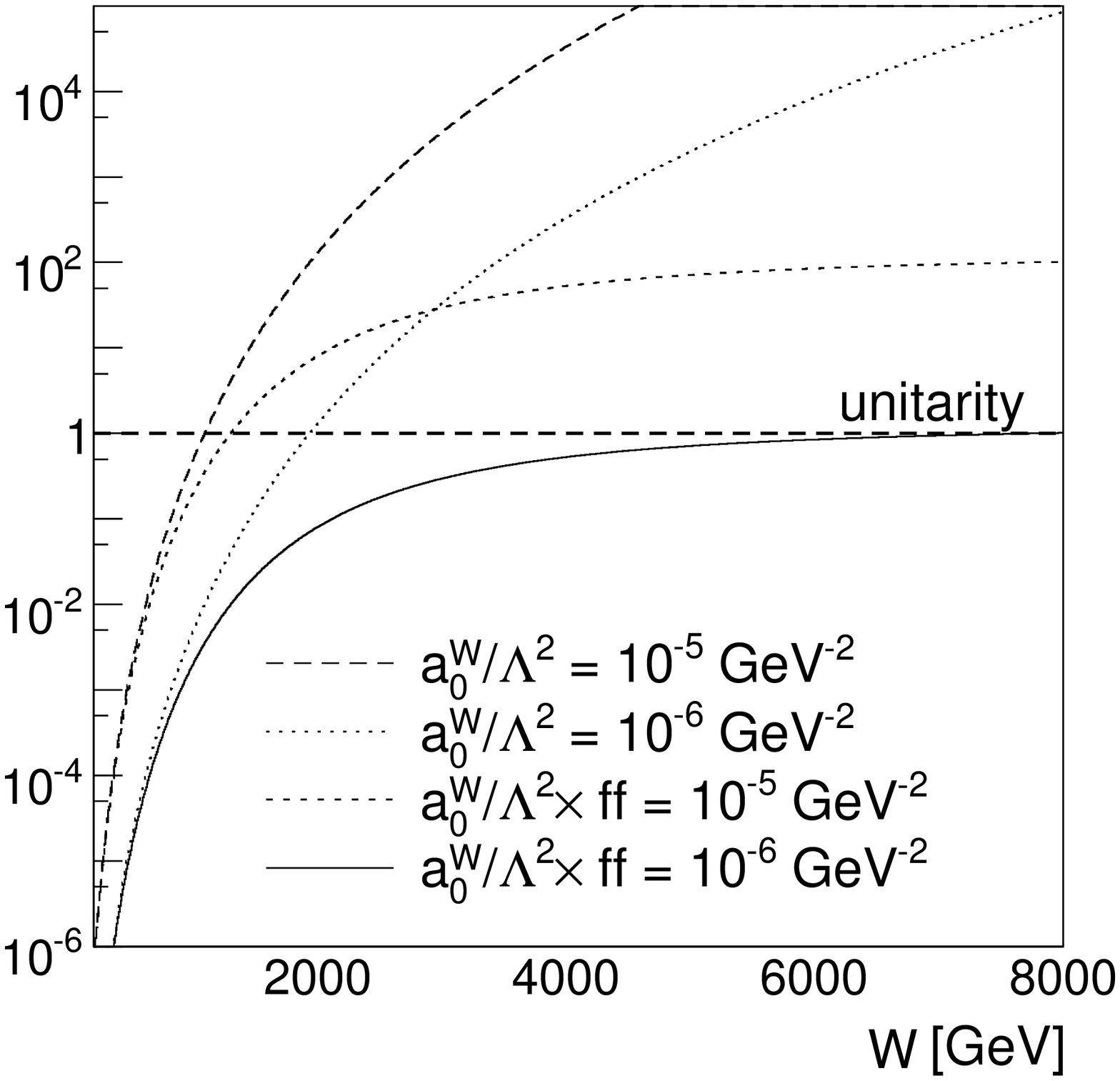}
\caption{
The unitarity condition \refeq{unitarityW} for quartic anomalous couplings $\aZerowLambda=10^{-5}$ and $10^{-6}\GeV^2$ as a function of two-photon invariant mass. The unitarity is violated above the horizontal line at 1. The form of the form factor \refeq{explicit_formfactor} is used for the two bottom curves.}
\label{fig:unitarity}
\end{figure}

The $WW$ and $ZZ$ two-photon cross sections rise quickly at high energies when 
any of the anomalous parameters are non-zero, as illustrated 
in \reffig{anom:totalxsectionQGC}. As it was already mentioned, the tree-level unitarity uniquely restricts the $WW\gamma\gamma$ coupling to the SM values at asymptotically high 
energies. This implies that any deviation of 
the anomalous parameters  $\aZerowLambda$, $\aCwLambda$, $\aZerozLambda$, $\aCzLambda$ from the SM zero value will eventually violate unitarity. 
Therefore, the cross section rise have to be 
regulated by a form factor which vanishes in the high energy limit to 
construct a realistic physical model of the BSM theory. At LEP where the 
center-of-mass energy was rather low, the wrong high-energy behavior did not 
violate unitarity; however, it must be reconsidered at the LHC. We therefore 
modify the couplings as introduced in \refeq{anom:lagrqgc} by form factors 
that have the desired behavior, i.e. they modify the coupling at small 
energies only slightly but suppress it  when the center-of-mass energy $W_{\gamma\gamma}$ 
increases. The form of the form factor that we consider
\be
a\rightarrow \frac{a}{(1+W^2_{\gamma\gamma}/\Lambda^2)^n}
\label{eq:anom:formfactor}
\ee
The exact form of the form factor is not imposed but rather only conventional 
and the same holds for the value of the exponent $n$.
$\Lambda^2$ corresponds to the scale where new physics should appear and where 
the new type of production 
would regularize the divergent high energy behavior of the 
Lagrangians \refeq{anom:lagrqgc}. 
\par The unitarity of the scattering $S$-matrix imposes a condition on the
partial waves amplitudes defined as 
\be
a_J(\sqrt{s})=\frac{1}{32}\int^1_{-1}\d(\cos\theta) \amp(\sqrt{s}, \cos\theta, 
a_0, a_C)P_J(\cos\theta) 
\ee
where $P_J(\cos\theta)$ are the Legendre polynomials depending on the polar 
angle in the $\gamma\gamma$ center-of-mass. The unitarity 
condition of the $J$ scattering amplitude in the $\gamma\gamma\rightarrow WW$ process reads 
\be
\beta \sum_{\lambda_1,\lambda_2}|a_J(\sqrt{s})|^2\le \frac{1}{4}
\ee
where $\beta=\sqrt{1-4m^2_W/s}$ is the velocity of a $W$ boson in the 
center-of-mass frame and the $\lambda_1,\,\lambda_2$ indices denote the
$W$ polarization states. 
\par For the anomalous interaction \refeq{anom:lagrqgc}, the most restrictive bounds come from the $J=0$
partial wave, which can be easily understood since $W$s with longitudinal polarizations without any spin flip are dominantly produced in this case. 
For $J=0$, the unitarity bounds read \cite{Eboli:2000ad}
\begin{eqnarray}
\frac{1}{N} \left(\frac{\alpha a s}{16}\right)^2\left(1-\frac{4M_W^2}{s}\right)^{1/2}
\left(3-\frac{s}{M^2_W}+\frac{s^2}{4M^4_W}\right)\leq 1\ \mathrm{for}\ V=W
\label{eq:unitarityW}\\
\frac{1}{N} \left(\frac{\alpha a s}{16\cos^2\theta_W}\right)^2\left(1-\frac{4M_Z^2}{s}\right)^{1/2}
\left(3-\frac{s}{M^2_Z}+\frac{s^2}{4M^4_Z}\right)\leq 1\ \mathrm{for}\ V=Z\label{eq:unitarityZ}
\end{eqnarray}
where $a=a_0/\Lambda^2$ or $a_C/\Lambda^2$ and $N=1/4$ $(4)$ for $a_0/\Lambda^2$ $(a_C/\Lambda^2)$.
\par
The unitarity violation in $\gamma\gamma\rightarrow WW$ process was 
investigated in the Ref.~\cite{Pierzchala:2008xc}. For relevant values of
$\aZerow$  which are to be probed at the LHC using forward detectors, it was found that the unitarity 
is violated around $W_{\gamma\gamma}=2\TeV$ for the form factor exponent 
$n=2$. We therefore adopt this type of form factor for the following study, 
i.e. the form factor
\be
a\rightarrow\frac{a}{\left[1+(W_{\gamma\gamma}/2\TeV)^2\right]^2}
\label{eq:explicit_formfactor}
\ee
is introduced for all quartic couplings $a=\aZerowLambda,\,\aZerozLambda,\,
\aCwLambda,\,\aCzLambda$.
The unitarity condition \refeq{unitarityW} for couplings $\aZerowLambda=10^{-5}$ and $10^{-6}\GeV^2$ is illustrated in \reffig{unitarity}. First we see that couplings without the form factors violate unitarity already at TeV energies. 
On the other hand, employing the form factors as described above justifies the non-violation of the unitarity of events inside the 
AFP acceptance ($W\lesssim 2\TeV$) if the resulting limits on neutral couplings $a_0$ are of the order of $10^{-6}\GeV^{-2}$.
For the charged couplings $a_C$ the unitarity condition is less strict due to  $N=4$ in Equations (\ref{eq:unitarityW}) and (\ref{eq:unitarityZ}).

%%%%%%%%%%%%%%%%%%%%%%%%%%%%%%%%%%%%%%%%%%%%%%%%%%%%%%%%%%%%%%%%%%
\subsection{Anomalous triple gauge $WW\gamma$ couplings}
%%%%%%%%%%%%%%%%%%%%%%%%%%%%%%%%%%%%%%%%%%%%%%%%%%%%%%%%%%%%%%%%%%

In this section, we discuss the implementation of the triple gauge $\wwgamma$ couplings
(TGC). The TGC have already been quite well 
constrained at LEP. The effective Lagrangian involving trilinear 
boson couplings with a photon will be introduced and used to study the 
sensitivities to the coupling parameters in two-photon events. Note that the 
lowest dimensional triple gauge boson operator $ZZ\gamma$ is of dimension six, 
the effect of this coupling in two-photon events will be the subject of a further study.
First, the effective Lagrangians describing the triple gauge couplings are 
introduced before evaluating the anomalous cross section.

%%%%%%%%%%%%%%%%%%%%%%%%%%%%%%%%%%%%%%%%%%%%%%%%%%%%%%%%%%%%%%%%%%
\subsubsection{Effective triple gauge boson operators}
%%%%%%%%%%%%%%%%%%%%%%%%%%%%%%%%%%%%%%%%%%%%%%%%%%%%%%%%%%%%%%%%%%

The most general form of an effective Lagrangian $\lumi_{WW\gamma}$ involving  
two charged vector bosons $W$ and one neutral vector boson has only seven 
terms which have the correct Lorentz structure (see \cite{Hagiwara:1986vm,
Kepka:2008yx} for 
details). 
This is because only seven out of the nine helicity states of the $W$ pair 
production can be reached with the spin-1 
vector boson exchange. The other two states have both $W$ spins pointing in 
the same direction with an overall spin 2.

Further more, only three out of the seven operators preserve the $P-,\ C-$ and 
$T-$ discrete symmetries separately.
We restrict ourselves to study this subset of operators. They are the 
following 
\be
   \lumi/g_{WW\gamma}=i(W^{+}_{\mu\nu}W^{\mu}A^{\nu}-W_{\mu\nu}W^{+\mu}A^{\nu})
   +i\kappa^{\gamma}W_{\mu}^{+}W_{\nu}A^{\mu\nu}+i\frac{\lambda^{\gamma}}{M_W^2}W^{+}_{\rho\mu}
   W^{\mu}_{\phantom{\mu}\nu}A^{\nu\rho}
\label{eq:anom:tgclag}
\ee
where the tensor is  $W_{\mu\nu}=\p_\mu W_\nu-\p_\nu W_\mu$, $g_{WW\gamma}=e$ 
is the trilinear coupling in the SM model whose strength is fixed by the 
charge of the $W$, and $\kappa^{\gamma}$ and $\lambda^{\gamma}$ are the 
anomalous parameters, and their values are 1 and 0 in the SM, respectively. 
They can be related to the magnetic $\mu_W$ and electric $Q_W$ moments of the 
$W^+$ by 
\begin{eqnarray}
\mu_W&=&\frac{e}{2m_W}(1+\dkap + \lam)\nonumber\\
Q_W&=&\frac{e}{m^2_W}(\dkap-\lam)
\end{eqnarray}
where $\dkap\equiv\kappa^{\gamma}-1$ describes the deviation of the parameter 
from the SM value.
(it is straightforward to verify that \refeq{anom:tgclag} gives the SM 
trilinear   
Lagrangian \refeq{anom:lagrww1} for $\kappa^{\gamma}=1$ and  
$\lambda^{\gamma}=0$. Our convention differs from the one 
in \cite{Hagiwara:1986vm} by a factor of -1).

The current best 95\% CL limits on anomalous couplings come from the combined 
fits of all LEP experiments~\cite{LEPlimits}.
\begin{equation}
   -0.098<\dkap<0.101 \quad -0.044<\lam<0.047
\end{equation}
The CDF collaboration presented the most stringent constraints on $\wwgamma$
coupling measured at hadron colliders~\cite{TEVlimits}  
\begin{equation}
   -0.51<\dkap<0.51 \qquad  -0.12<\lam<0.13
\end{equation}
analyzing the $W\gamma$ events in parton-parton interactions.
Even though the LEP results are more precise than the results from the hadron 
collider, there is always a mixture of $\gamma$ and $Z$ exchanges present in the
process $e^+e^-\rightarrow WW$ from which the couplings are extracted. The 
two-photon $WW$ production at the LHC has the advantage that pure $W-\gamma$ couplings are tested
and no SM $Z$ exchange is present.

%%%%%%%%%%%%%%%%%%%%%%%%%%%%%%%%%%%%%%%%%%%%%%%%%%%%%%%%%%%%%%%%%%
\subsubsection{Anomalous cross section}
\begin{figure}
\centering
\includegraphics[width=\twopicwidth]{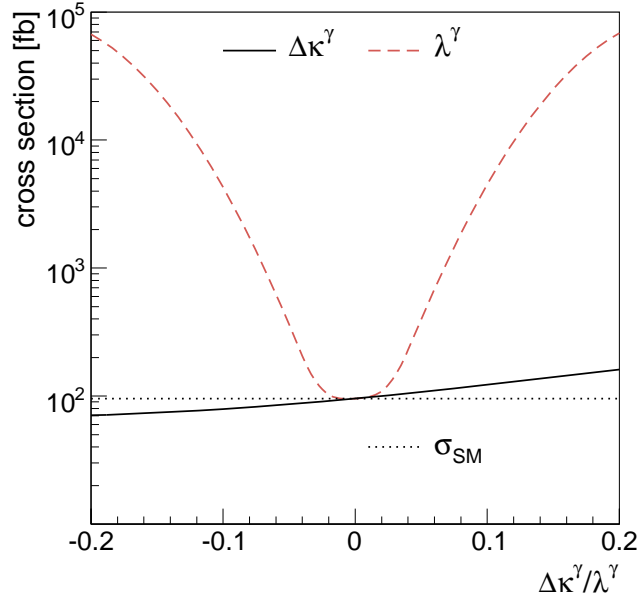}
\caption{Enhancement of the total cross section with the triple-boson 
anomalous couplings $\dkap$, $\lam$. 
The rise of the cross section due to $\lam$ is well pronounced whereas the 
dependence on $\dkap$ is modest (the tail for 
large negative $\dkap$ where cross section increases is not shown).}
\label{fig:anom:totalxsectionTGC}
\end{figure}

\begin{figure}
\centering
\includegraphics[width=\twopicwidth]{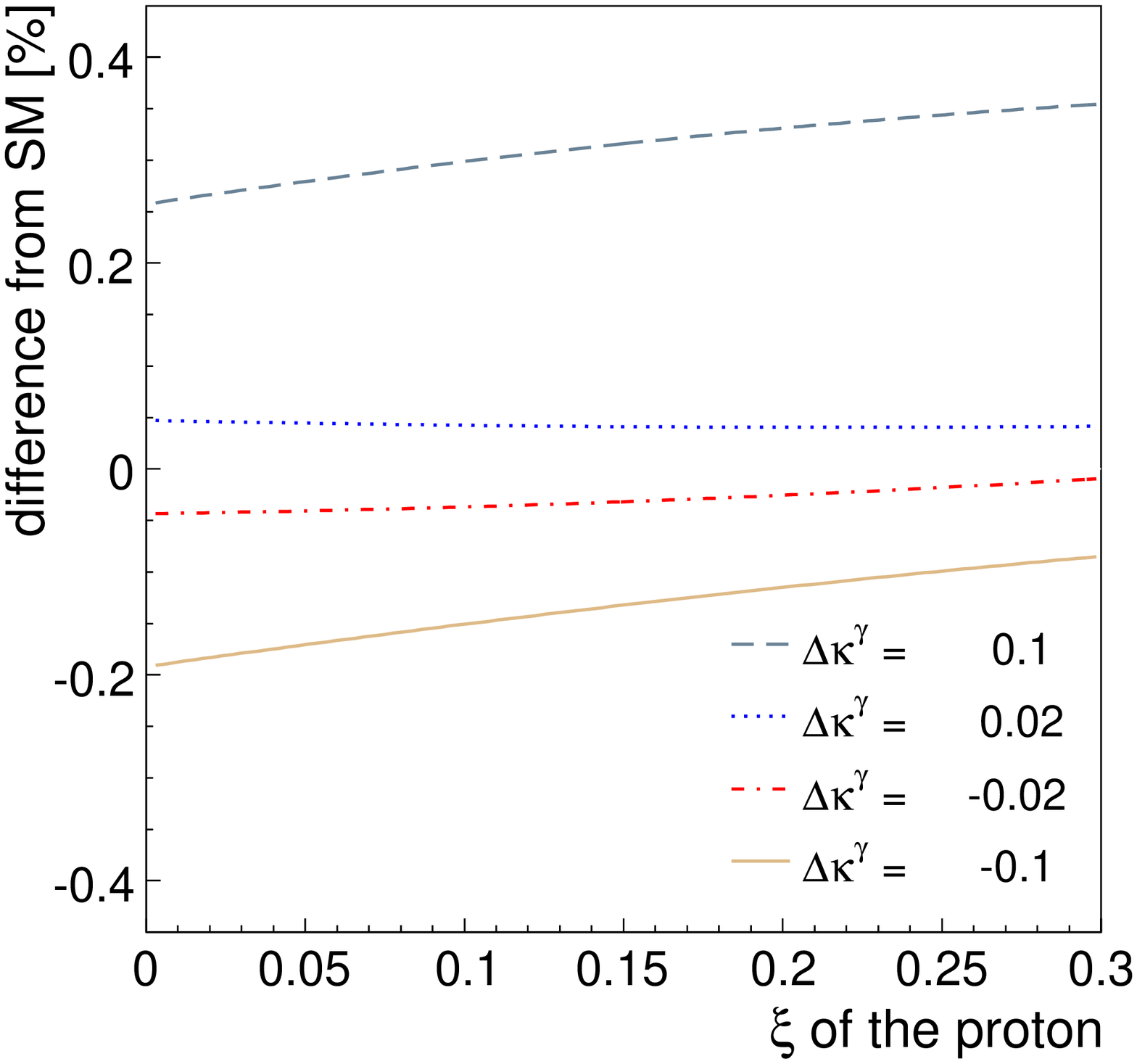}
\includegraphics[width=\twopicwidth]{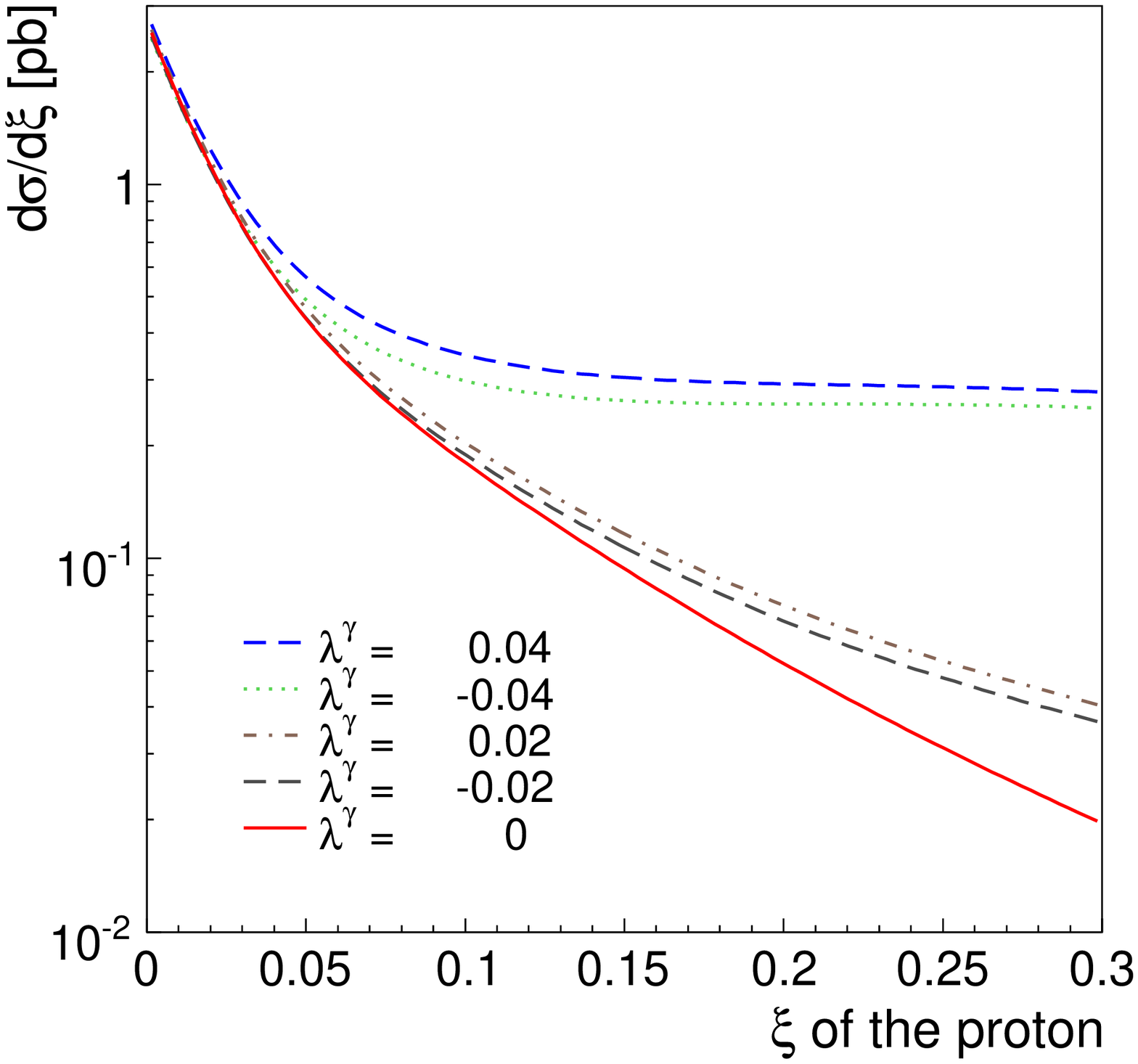}
\caption{$\xi$ dependence of the two-photon $WW$ cross section for different 
values of 
$\dkap$ (left) and $\lam$ (right) (SM values are 0).
For $\lam$, the cross section is enhanced at high 
$\xi$ which is at the edge of the forward detector acceptance ($\xi=0.15$). 
On the contrary, varying $\dkap$ in the interesting 
range $(-0.05<\dkap<0.05)$ changes mainly the normalization and not the shape 
of the $\xi$ distribution.}
\label{fig:anom:triple_xi}
\end{figure}

%%%%%%%%%%%%%%%%%%%%%%%%%%%%%%%%%%%%%%%%%%%%%%%%%%%%%%%%%%%%%%%%%%
The effect of the two anomalous couplings is different. The total cross 
section is much more sensitive to the anomalous 
coupling $\lam$. As shown in \reffig{anom:totalxsectionTGC}, the SM cross 
section $\sigma_{SM}=95.6\fb$
is a global minimum with respect to the $\lam$ parameter. For $\dkap$ the 
minimum also exists but for large negative values
which have already been excluded by experiments. The last term proportional 
to $\lam$ in \refeq{anom:tgclag} does not 
have a dimensionless coupling. With simple dimensional consideration we see 
that the $\gamma\gamma\rightarrow WW$ scattering amplitude which has to be 
dimensionless will have the form $\sim \frac{W^4}{M^4_W}$ and will therefore 
be quickly rising 
as a function of the two-photon mass $W$. This is seen 
in \reffig{anom:triple_xi} where the cross section is 
shown as a function of the momentum fraction loss of the proton. $\dkap$ 
enhances the overall normalization of the distribution (left) whereas $\lam$ 
gives rise to the $\xi$ tail (right) as anticipated. 
\subsection{Coupling form factors}
The rise of the cross section for anomalous TGC at high energy leads again to 
the violation of unitarity. The enhancement of the cross section has to be 
again regulated by appropriate from factors. We apply the same form factors as 
already mentioned for the quartic couplings \refeq{explicit_formfactor}. 
%

%%%%%%%%%%%%%%%%%%%%%%%%%%%%%%%%%%%%%%%%%%%%%%%%%%%%%%%%%%%%%%%%%%
\section{The Forward Physics Monte Carlo}
%%%%%%%%%%%%%%%%%%%%%%%%%%%%%%%%%%%%%%%%%%%%%%%%%%%%%%%%%%%%%%%%%%

In this section, we briefly describe the Forward Physics Monte Carlo (FPMC) 
generator~\cite{fpmc} used
extensively in this paper to produce all signal and background events.
FPMC aims to accommodate all relevant models for  
forward physics which could be studied at the LHC and contains in particular
the two-photon and double pomeron exchange processes which 
are relevant for this study since we focus on events in which both protons are detected. The generation of the forward processes is embedded 
inside HERWIG~\cite{herwig}. The advantage of the program is that all the processes 
with leading protons can be studied in the same framework, using the same 
hadronization model. It is dedicated to generate the following exchanges:
\bi
\item two-photon exchange
\item single diffraction
\item double pomeron exchange
\item central exclusive production
\ei

In FPMC, the diffractive and exclusive processes are implemented by modifying 
the HERWIG routine for the 
$e^+e^-\rightarrow(\gamma\gamma)\rightarrow X$ process. In case of the 
two-photon $pp$ events, as we mentioned in Section I,
the Weizs\"{a}cker-Williams (WWA) formula describing 
the photon emission off point-like electrons is substituted by the Budnev
flux~\cite{Budnev} 
which describes properly the coupling of the photon to the proton, taking into 
account the proton electromagnetic structure. 
\par
The effective Lagrangians parametrizing new interactions of electroweak bosons mentioned explicitly in Equations \ref{eq:anom:lagrqgc} and \ref{eq:anom:tgclag} are functions of six anomalous parameters: $\dkap$, $\lam$ for the triple gauge couplings and $\aZerowLambda,\,\aZerozLambda,\,\aCwLambda,\,\aCzLambda$ for the quartic ones. The corresponding matrix elements squared were obtained with the CompHEP program \cite{comphep} whose output was interfaced with FPMC.
\par
The single diffractive and double pomeron exchange events are produced in FPMC
using the diffractive parton densities measured at HERA~\cite{herapdf}.  The 
outcome of the QCD Dokshitzer-Gribov-Lipatov-Altarelli-Parisi~\cite{dglap}
fits to the proton diffractive structure functions are the values of the 
pomeron and reggeon trajectories 
$\alpha_\pom(t)=\alpha_\pom(0)+t\alpha'_\pom$, $\alpha_\reg(t)=
\alpha_\reg(0)+t\alpha'_\reg$ governing the corresponding flux energy and $t$ 
dependences, and the pomeron parton distribution 
functions.

In addition, due to the factorization breaking between LHC and HERA, an
additional survival probability~\cite{survival} is introduced and it is assumed
to be 0.03 for DPE and 0.9 for photon exchanges in the following. Technically,
in FPMC, for processes in which the partonic structure of the pomeron is probed,  
the existing HERWIG matrix elements of non-diffractive production are used to 
calculate the production cross sections. The list of particles is corrected 
at the end of each event to change the type of particles from the initial 
state electrons to hadrons and from the exchanged photons to 
pomerons/reggeons, or gluons, depending on the process. 

The output of the FPMC generator was interfaced with the fast 
simulation of the ATLAS detector in the standalone ATLFast++ package for 
ROOT~\cite{atlfast}. The fast simulation of ATLAS is performed for all signal
and background processes.

%
%%%%%%%%%%%%%%%%%%%%%%%%%%%%%%%%%%%%%%%%%%%%%%%%%%%%%%%%%%%%%%%%%%
\section{Selection of diffractive and photon exchange events
at high luminosity at the LHC}
\label{sec:selection}
%%%%%%%%%%%%%%%%%%%%%%%%%%%%%%%%%%%%%%%%%%%%%%%%%%%%%%%%%%%%%%%%%%

In this section, we detail briefly the methods used to select diffractive and
two-photon exchange events at the LHC in the ATLAS detector. The same study
could be made using the CMS detector which would lead to similar results. 
At high instantaneous luminosity at the LHC, it is not 
possible to
use the so-called standard rapidity gap method since up to 
30 interactions 
--- one hard interaction and many minimum bias events --- occur in the
same bunch crossing. The $WW$ exclusive production overlaps with soft
interactions which fill the gap devoid of any energy and the gap selection does
not work any longer. 

At high
luminosity, the only method to select the diffractive and photon exchange events 
is to detect the intact protons in the final state. We thus assume the existence
of forward proton detectors in the ATLAS (or CMS) detectors. A project
called AFP (ATLAS Forward Physics) is under evaluation in the ATLAS
collaboration and corresponds to the installation of forward detectors at 220
and 420 m allowing to detect intact protons in the final state~\cite{afp}. The acceptance
of such detectors is about 0.0015 $<\xi <$ 0.15 where $\xi$ is the proton momentum
fraction carried by the pomeron or the photon.

%%%%%%%%%%%%%%%%%%%%%%%%%%%%%%%%%%%%%%%%%%%%%%%%%%%%%%%%%%%%%%%%%%
\section{Measuring the $pp\rightarrow pWW\!p$ process in the Standard Model}
%%%%%%%%%%%%%%%%%%%%%%%%%%%%%%%%%%%%%%%%%%%%%%%%%%%%%%%%%%%%%%%%%%

Before discussing the possibility of observing anomalous couplings, we
will mention how to discover the SM $pWW\!p$ process at the LHC. 

%%%%%%%%%%%%%%%%%%%%%%%%%%%%%%%%%%%%%%%%%%%%%%%%%%%%%%%%%%%%%%%%%%
\subsection{The $pp\rightarrow pWW\!p$ signal}
%%%%%%%%%%%%%%%%%%%%%%%%%%%%%%%%%%%%%%%%%%%%%%%%%%%%%%%%%%%%%%%%%%

\begin{figure}
\centering
\includegraphics[width=\picwidthsq]{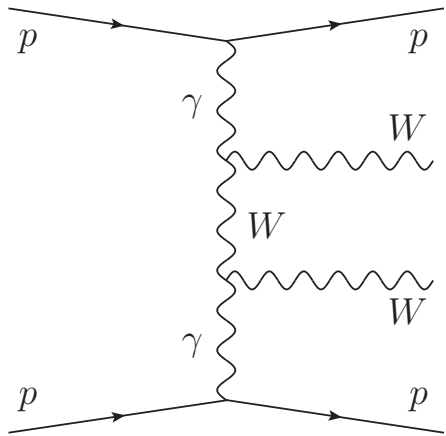}
\caption{Diboson production through the two-photon exchange. Intact protons 
leave the interaction scattered at small angles
$\lesssim20\units{mrad}$.}
\label{fig:anom:wwinpp}
\end{figure}

The total cross section of the exclusive process $pp\rightarrow pWW\!p$ where 
the interaction proceeds through the exchange of
two quasi-real photons shown in \reffig{anom:wwinpp} is 95.6\fb{} and this value 
has to be corrected for the survival probability factor 0.9.
\par
The cross section is rather modest in comparison to the inelastic production 
which is about three orders of magnitude higher (at $\sqrt{s}=14\TeV$, the 
NLO $W^+W^-$ cross section is 111.6\pb{}, produced via quark-anti-quark 
annihilation $q\bar{q}\rightarrow W^+W^-$ ($\sim95\%$) and also via 
gluon-gluon fusion $gg\rightarrow W^+W^-$ ($\sim5\%$)). A substantial amount 
of luminosity has therefore to be collected to have a significant $WW$ sample. 
It can only be accumulated when running at high LHC instantaneous 
luminosities $\lumi=10^{33}-10^{34}$\lumiunit. 
Under such running conditions, the two-photon events must
be selected with the forward proton tagging detectors. 
\par
The $W$ boson decays hadronically ($\sim68\%$) or leptonically ($\sim 32\%$). 
The hadronic or semi-leptonic decays in which at least one jet is present 
could be mimicked by the QCD dijets or non-diffractive $WW$ production, 
overlaid
with other minimum bias interactions leading to a proton hit in the forward 
detectors. For simplicity, we focus on the $W$ decays only into electrons or 
muons in the final state. This in turn means that also only the leptonic 
decays of the $\tau$ lepton  ($\sim35\%$) are considered. Semi-leptonic decays
of the $W$s will be considered in a further study. About $\sim6\%$ of 
the total $WW$ cross section is retained for the analysis.
About 1800 events are produced with two leptons in the final states for 
30\invfb, an integrated luminosity which corresponds approximately to the 3
first
years of running. We will see further that taking into account the forward 
detector acceptance, and the electron/muon reconstruction efficiencies, 
the expected number of events
drops down to 50 events.

%%%%%%%%%%%%%%%%%%%%%%%%%%%%%%%%%%%%%%%%%%%%%%%%%%%%%%%%%%%%%%%%%%
\subsection{Diffractive and $\gamma\gamma$ dilepton background}
%%%%%%%%%%%%%%%%%%%%%%%%%%%%%%%%%%%%%%%%%%%%%%%%%%%%%%%%%%%%%%%%%%

\begin{figure}
\includegraphics[width=\twopicwidth]{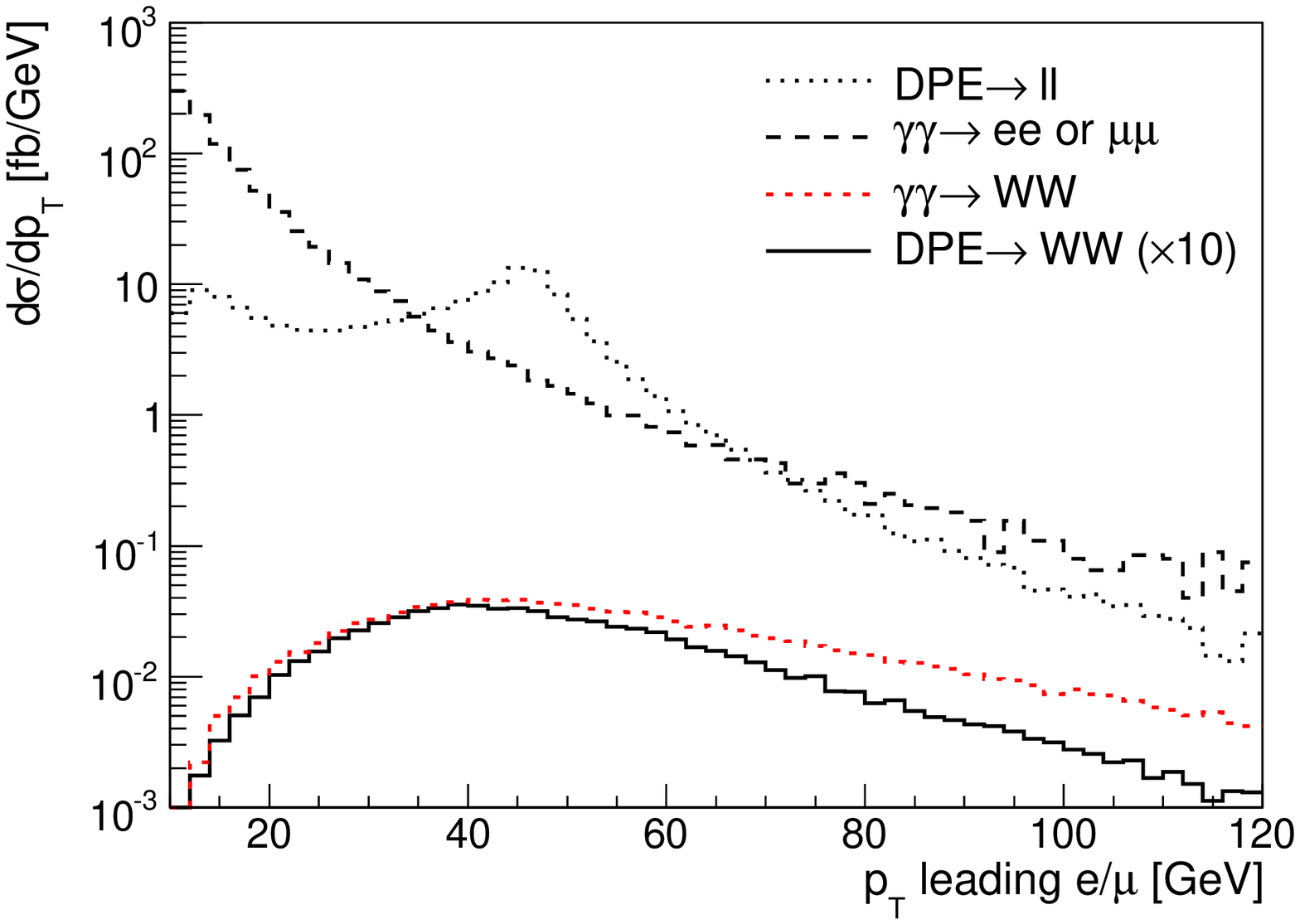}
\includegraphics[width=\twopicwidth]{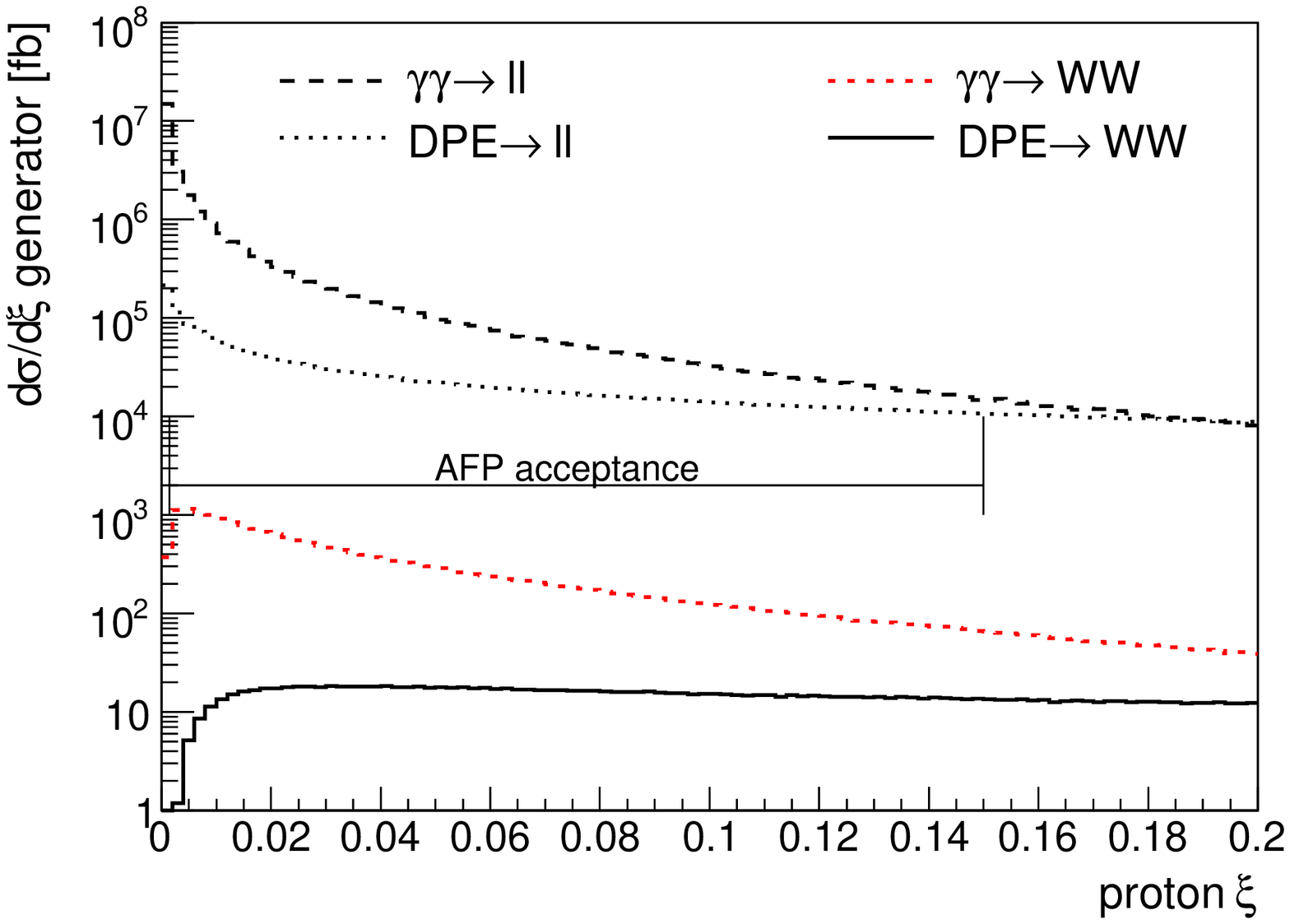}
\caption{Transverse momentum of the leading $e$ or $\mu$ (left) and the 
momentum fraction loss $\xi$ (right) distributions
for processes which have two leptons as well as two forward intact protons in 
the final state. The signal 
$\gamma\gamma\rightarrow WW$ is outraged by the $ll$ two-photon and DPE 
production.}
\label{fig:anom:withoutcuts}
\end{figure}
The clean two-leptonic signature of the two boson signal process
$\gamma\gamma\rightarrow W^{+}W^{-} \rightarrow l\bar{l}\nu\bar{\nu} $
can be mimicked by several background processes which all have
two intact protons in the final state. They are the following:
\begin{enumerate}
\item $\gamma\gamma\rightarrow l\bar{l}$ - two-photon dilepton production 
\item DPE$\rightarrow l\bar{l}$ - dilepton production through double pomeron 
exchange
\item DPE$\rightarrow W^+W^-\rightarrow l\bar{l}\nu\bar{\nu}$ - diboson 
production through double pomeron exchange
\end{enumerate}

The Double Pomeron Exchange (DPE) production of dileptons and dibosons is 
described within the factorized Ingelman-Schlein model where the hard 
diffractive scattering is interpreted in terms of the
colorless pomeron with a partonic structure. Cross sections are obtained as a 
convolution of the hard matrix elements with 
the diffractive parton density functions measured at HERA~\cite{herapdf}. 
Dileptons in DPE are produced as Drell-Yan pairs, probing the quark structure 
of the pomerons. The exchange is carried
out through $\gamma^*$ or $Z^*$. Contrary to the two-photon exclusive case 
where only scattered protons and leptons in the central detector are present, 
in DPE events, pomeron remnants accompany the interacting partons. They give a 
significant boost to the 
lepton pair in the transverse plane resulting in a non-negligible azimuthal 
decorrelation $\Delta\phi$ between the leptons.
 Finally, the diboson production in DPE is very similar to the actual 
 $\gamma\gamma\rightarrow WW$ signal except that the 
mass distribution of the $WW$ system is not as strongly peaked towards small 
values. The DPE dilepton and diboson total 
production cross sections at generator level are respectively 743\pb{} 
(all lepton families) and 211\fb{} (all decay modes).

As we already mentioned, the experimental signature of the two-photon or DPE 
interaction in which two scattered protons go intact in the 
beam pipe and can be tracked in forward detectors can be lost by additional 
soft interactions between the outgoing
protons. 
The survival probabilities for the 
QED two-photon processes and QCD diffractive and central exclusive processes 
are respectively taken to be 0.9 and 0.03~\cite{survival}. The 
mentioned cross sections have to be therefore 
multiplied by these survival probability factors yielding cross sections of 
the signal and background shown in \reftab{anom:totalxsection}. The dilepton 
production is the largest background, three orders of magnitude higher than 
the desired $\gamma\gamma\rightarrow WW$ signal.
\begin{table}
\centering 
\begin{tabular}{|c|c|}
\hline
 process   & total cross section   \\
\hline
$\gamma\gamma\rightarrow WW $ &  96.5\fb   \\
$\gamma\gamma\rightarrow ll$ $(p_T^{lep1}>5\GeV)$ &  39.4\pb  \\
DPE$\rightarrow ll$    &  7.4\pb  \\
DPE$\rightarrow WW$   & 8.1\fb  \\
\hline
\end{tabular}
\caption{Total cross sections for SM $\gamma\gamma\rightarrow WW$ signal and 
background processes at 14\TeV{} including the gap survival probability 
factor (0.9 for QED and 0.03 for DPE processes).}
\label{tab:anom:totalxsection}
\end{table}

The characteristic properties of the two-photon and DPE productions are 
visible in~\reffig{anom:withoutcuts}. The leptons ($e/\mu$) are required 
to be within the generic central detector acceptance $p_T^{lep1,2}>10\GeV$, 
$|\eta^{lep}|<2.5$. The \pt{} distributions (left) are peaked towards 0. Since 
the leptons are predominantly produced at central pseudo-rapidity this 
reflects the steepness of the  two-photon luminosity  dependence as a function 
of $W_{\gamma\gamma}$. In the DPE dilepton spectrum
one can identify the $Z^*$ resonance around $p_T^{lep1}=45\GeV$. The diboson 
spectrum on the other hand slowly increases until the
$WW$ channel is totally kinematically opened and then decreases due to the drop 
of the effective photon-photon or pomeron-pomeron luminosity. On the right 
side of \reffig{anom:withoutcuts}, the momentum fraction loss $\xi$ 
distribution  shows again that the two-photon production is dominant at low 
mass. The momentum fraction tail of the DPE is truncated at $\xi=0.2$ which is 
about the limit of 
the validity of the factorized pomeron model. The acceptance of the AFP 
detectors is shown as well. It provides us an access of two-photon masses up 
to $\sqrt{s}\times\xi_{max}=14\TeV\times0.15=2.1\TeV$.

\begin{figure}
\includegraphics[width=\twopicwidth]{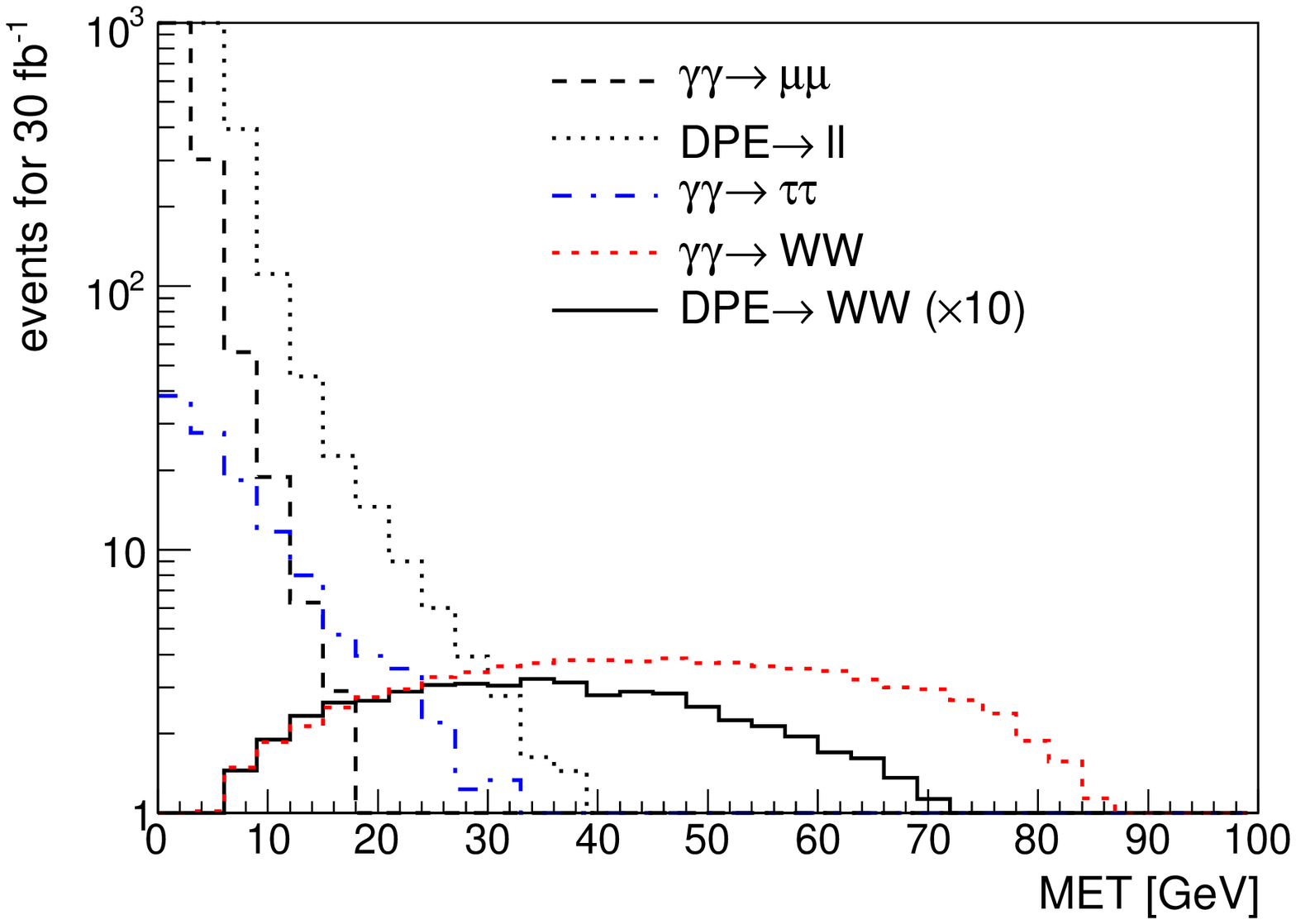}
\includegraphics[width=\twopicwidth]{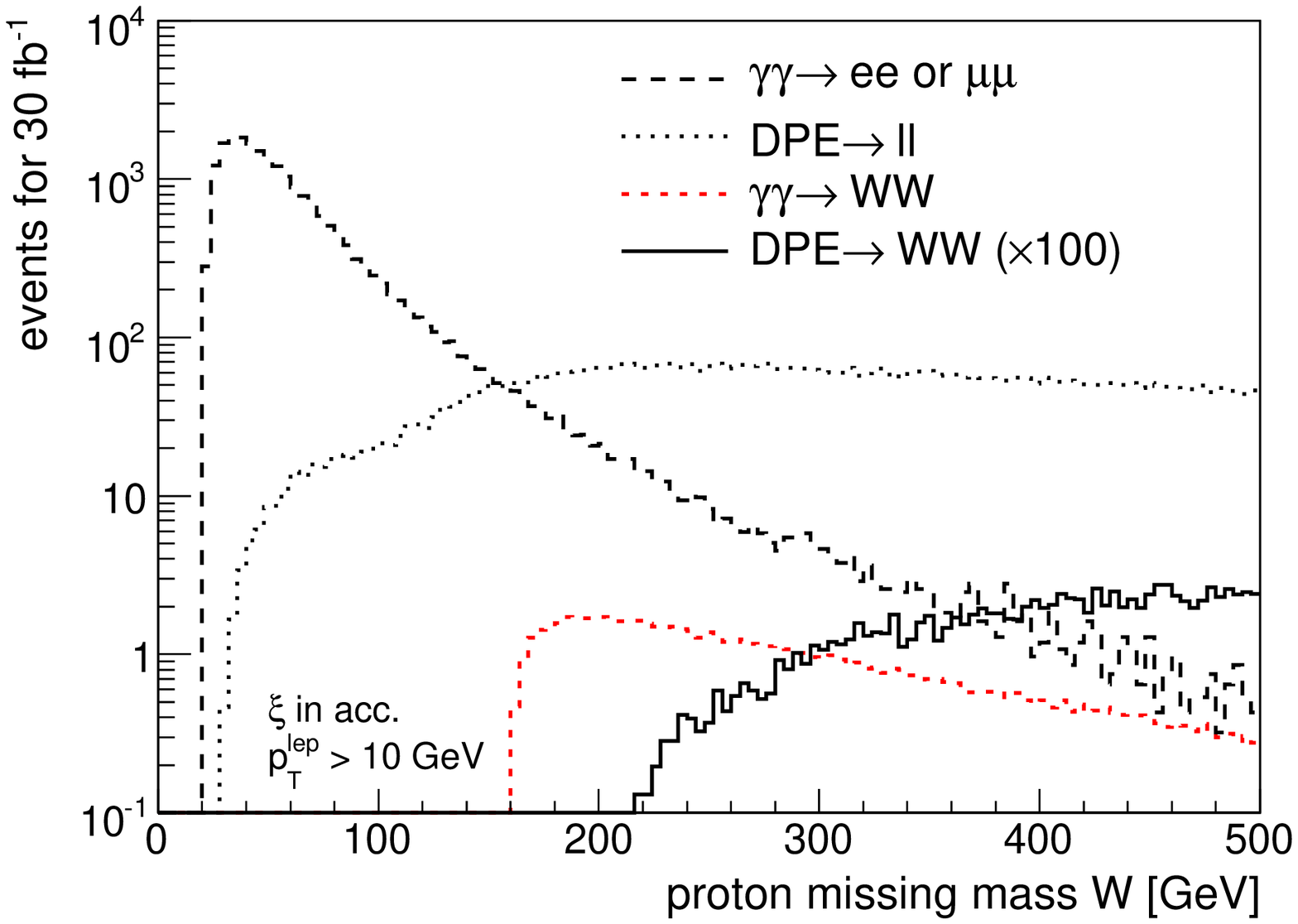}
\caption{Missing transverse $\missET$ energy (left) and reconstructed $W$ 
missing mass  in the forward detectors (right) for the two-photon $WW$ signal 
and background processes. The $WW$ signal has a production threshold at 
$2m_W$ and has a large $\missET$
due to the undetected neutrinos.}
\label{fig:anom:met_and_w}
\end{figure}

The most natural distinction of the diboson signal is the missing transverse 
energy (\missET) in the event due to the undetected two neutrinos, 
see  \reffig{anom:met_and_w} (left). It provides a very effective suppression 
not only of the two-photon dileptons where leptons are produced back-to-back 
in the central detector with no intrinsic \missET{}, but suppresses also the 
DPE dilepton background, even though  some of the energy is lost due to the pomeron remnants 
is not seen in the calorimeter. It can be due to either
a limited $\eta$ coverage of the calorimeter or due to a minimum energy 
readout threshold in the system which the pomeron remnants do not pass.  Both 
cases mimic \missET.

\begin{figure}
\centering
\includegraphics[width=\twopicwidth]{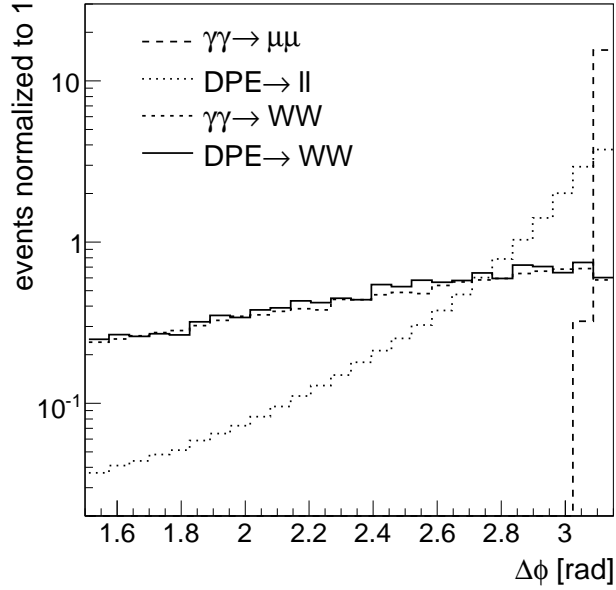}
\caption{$\Delta\phi$ between two leading leptons.  Dilepton events are more 
back-to-back than diboson events. DPE dileptons
is less peaked because of the presence of the pomeron remnants which gives a 
transverse boost to the Drell-Yan system.}
\label{fig:anom:dphi_ww}
\end{figure}

Another way to distinguish the diboson signal is to use the missing mass 
$W=\sqrt{\xi_1 \xi_2s}$ reconstructed in forward detectors which is shown 
in \reffig{anom:met_and_w} (right). The dilepton production is dominant at low 
mass in both 
two-photon and DPE exchanges, but has also a non-negligible contribution at 
high mass. The azimuthal angle $\Delta\phi$ between  the two leading leptons 
is depicted in \reffig{anom:dphi_ww}. Dilepton events are more back-to-back 
than the diboson ones.

As mentioned before, all signal and background processes are generated using 
FPMC, interfaced with the fast 
simulation of the ATLAS detector in the standalone ATLFast++ package. The aim
was to examine the general properties of all backgrounds in a fast way to 
define the strategies for early data measurements with the emphasis on the 
two-photon dilepton and anomalous coupling studies. 
Effects of the charge or jet mis-identifications cannot be considered in this
study using a fast simulation of the ATLAS detector but 
will be evaluated with real data. 
 
We will now discuss how to select the signal $\gamma\gamma\rightarrow WW$ 
events from the mentioned background.

%%%%%%%%%%%%%%%%%%%%%%%%%%%%%%%%%%%%%%%%%%%%%%%%%%%%%%%%%%%%%%%%%%
\subsection{Strategy to measure the  $pp\rightarrow pWW\!p$ process}
%%%%%%%%%%%%%%%%%%%%%%%%%%%%%%%%%%%%%%%%%%%%%%%%%%%%%%%%%%%%%%%%%%
\label{kap:anom:measurement}

It is necessary to use forward detectors to search for $pp\rightarrow pWW\!p$ 
production at high luminosity. 
After tagging the protons with a momentum fraction $0.0015<\xi_{1,2}<0.15$, 
the signal is selected with $\missET>20\GeV$ measured in the central detector 
and a missing mass $W>2m_W$ measured in 
forward proton detectors (computed as $\sqrt{\xi_1 \xi_2 s}$ where
$\xi_{1,2}$ and $\sqrt{s}$ are the proton momentum fraction loss and
the center-of-mass energy, respectively). Both cuts are natural for diboson production. 
The $\gamma\gamma\rightarrow ll$ production where leptons are produced 
back-to-back is completely removed requesting the azimuthal angle between the 
two observed leptons $\Delta\phi<2.7\rad$.

Let us note in addition that triggering on those events is quite easy since we
have two $W$s in the central ATLAS detectors decaying into leptons. 
The trigger menus of ATLAS are designed in a way to have the least possible 
prescales on leptons produced in electroweak bosons $W/Z$ decays. The L1 and 
High Level Triggers (HLT) can be operated without prescales
up to luminosities $\lumi=2\times10^{33}\lumiunit{}$ with  thresholds of 
20\GeV{} for single muons, and 18\GeV{} at L1 and 22\GeV{} at the HLT for 
single electrons~\cite{Aad:2009wy}. For higher luminosities, the trigger menus will have to be 
studied and tuned. In addition, most of the protons will be detected in the
forward proton detectors located at 220 m which can give an additional L1
trigger.

\begin{figure}
\includegraphics[width=\twopicwidth]{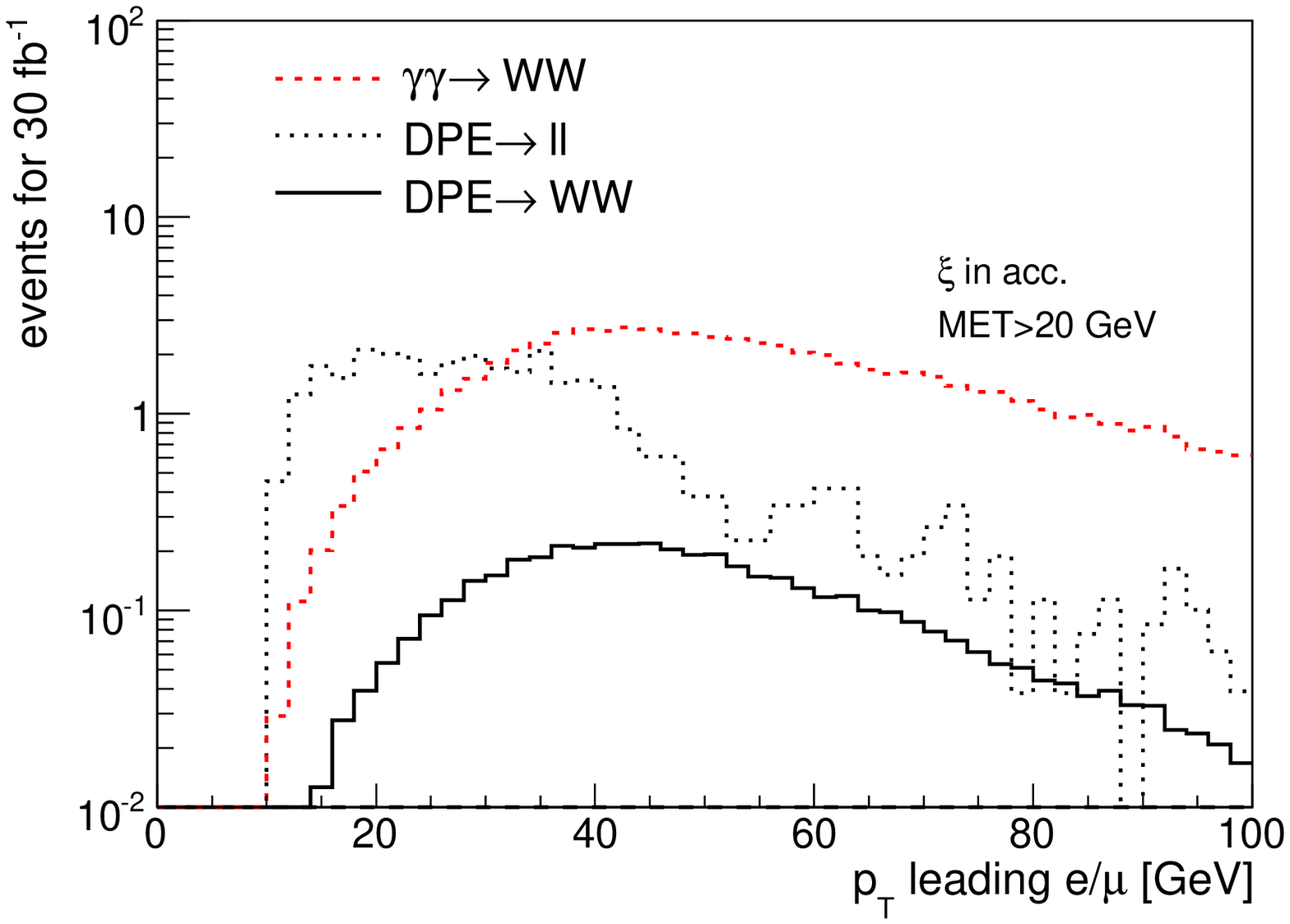}
\includegraphics[width=\twopicwidth]{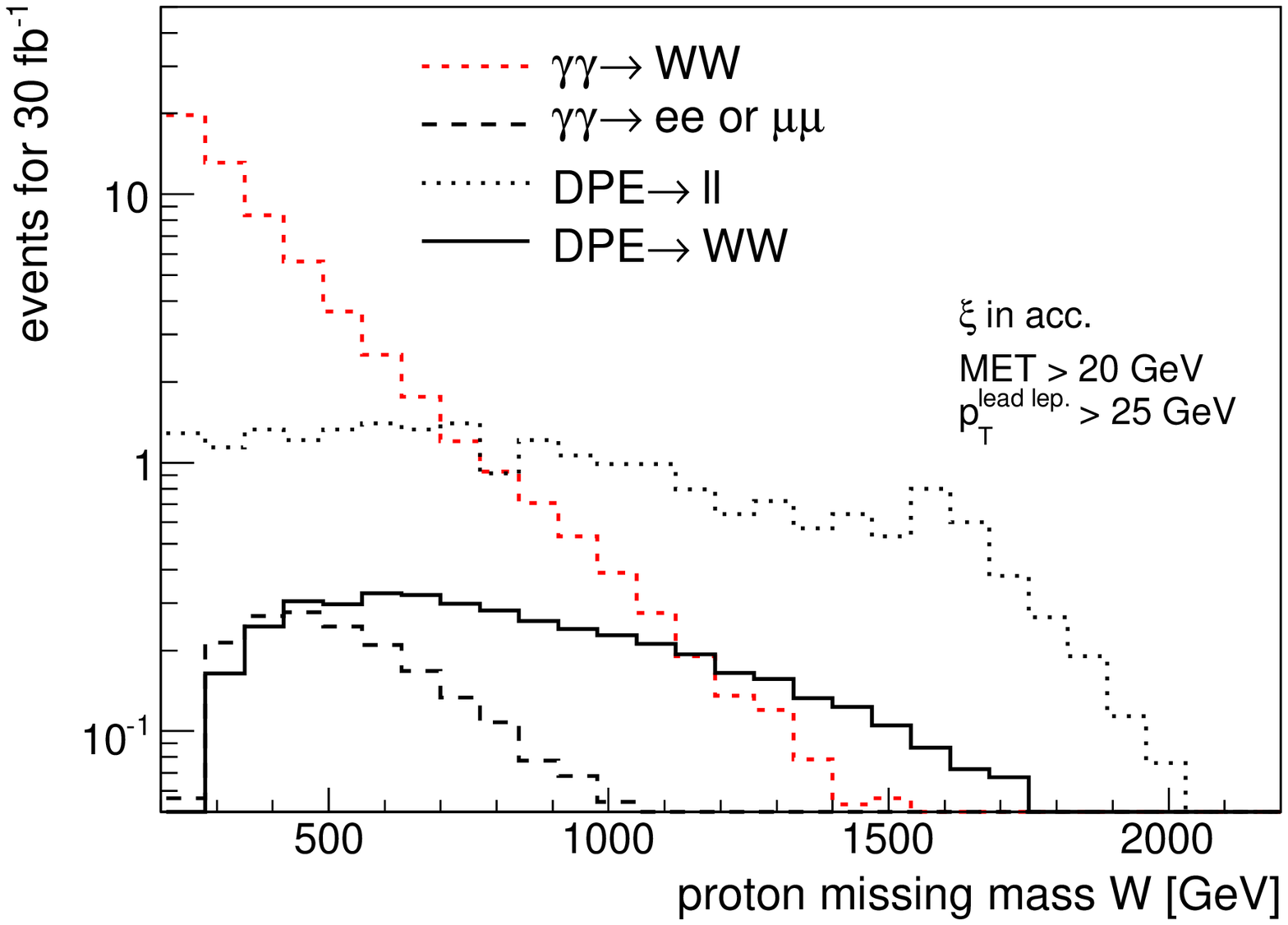}
\caption{Signal $\gamma\gamma\rightarrow WW$ and background before the cut on 
the leading lepton transverse momentum $\pt{}>$25\GeV{} (left) and before the 
cut on the missing mass $W<500\GeV$ (right). Both constraints are aimed 
to suppress the DPE$\rightarrow ll$ production which is the most 
important background for the measurement.}
\label{fig:anom:pt_and_w}
\end{figure}

The remaining background is composed of the DPE$\rightarrow ll$ ($\sim 80\%$) 
and DPE $\rightarrow WW$ (20\%). We handle it by requesting the transverse 
momentum of the leading lepton $p_T^{lep1}>25\GeV$  and the missing mass 
smaller than $W<500\GeV$, see \reffig{anom:pt_and_w}. This leaves us with the 
cross section $1.69\pm0.01\fb{}$ for the total background (the shown 
uncertainty reflects the statistical uncertainty of the calculation).  In 
summary, the following requirements are used:
\be
p_T^{lep1}>25\GeV,\,
 p^{lep2}_T>10\GeV,\,	      
 0.0015<\xi<0.15,\,	     
 \missET>20\GeV,\,    
 160<W<500\GeV,\,      
 \Delta\phi<2.7\rad
\ee
\begin{table}
\begin{tabular}{|c|ccc|cc||c|}
%\multicolumn{7}{c}{events for 30\invfb} \\
\hline
cut / process &  $\gamma\gamma\rightarrow ee$ & 
$\gamma\gamma\rightarrow \mu\mu$ & $\gamma\gamma\rightarrow \tau\tau$  & 
DPE$\rightarrow ll$ & DPE$\rightarrow WW$ & $\gamma\gamma\rightarrow WW$\\
\hline
\hline
gen. $p^{lep1}_T>5\GeV$ & 364500 & 364500   &  337500  & 295200  & 530 & 1198 \\
$p^{lep1,2}_T>10\GeV$	       & 24896 &        25547&        177&        17931&         8.8&        95 \\           
$0.0015<\xi<0.15$	       & 10398 &       10535 &       126 &       11487 &       5.9 &       89 \\       
$\missET>20\GeV$              &    $0$ &        0.86 &       14  &        33 &       4.7 &      78 \\
$W>160\GeV$       &         0 &                 0.86  &       8.3 &       33 &        4.7 &     78 \\   
$\Delta\phi<2.7$    &      0 &             0 &             0 &       14 &       3.8 &        61 \\       
$p_T^{lep}>25\GeV$   &      0 &             0 &             0 &        7.5 &        3.5 &        58 \\   
\hline   
$W<500$	        &      0 &             0 &             0 &       1.0 &      0.67 &       51  \\      
$\xi<0.1$ &   0 &             0 &             0 &      0.85 &      0.54 &       47  \\     
$\xi<0.05$&    0 &             0 &             0 &      0.40 &      0.25 &       32  \\
\hline
\end{tabular}
\caption{Background rejection to select $\gamma\gamma\rightarrow WW$ events 
for \lumi=30\invfb. The overall final signal is 51,
47, 32 signal events for the upper limit of the forward detector acceptance 
$\xi_{max}=0.15$, 0.1, and 0.05, respectively, whereas the background is as 
low as 1.7, 1.4, 0.65 events. The statistical uncertainty on the expected 
number of events is at most 15\% and is the largest for DPE$\rightarrow ll$. 
}
\label{tab:anom:measureww_cuts}
\end{table}
The successive effects of all mentioned constraints are given 
in \reftab{anom:measureww_cuts} where the number of events is shown for 
30\invfb. In three years, one expects about $50.8\pm0.2$ signal events and 
$1.7\pm0.1$ background events. 
It is interesting to notice that this measurement can be successfully carried 
out even if the AFP acceptance does not 
reach its design maximum acceptance range $\xi_{max}=0.15$. 
The number of expected events for $\xi_{max}=0.1$, and $\xi_{max}=0.05$ 
are $47\pm0.2$, $32\pm0.2$ for 30\invfb. The corresponding total backgrounds 
are $1.5\pm0.1$ and $0.74\pm0.08$, 
respectively.

%%%%%%%%%%%%%%%%%%%%%%%%%%%%%%%%%%%%%%%%%%%%%%%%%%%%%%%%%%%%%%%%%%
\subsection{Results}
%%%%%%%%%%%%%%%%%%%%%%%%%%%%%%%%%%%%%%%%%%%%%%%%%%%%%%%%%%%%%%%%%%
The $5\sigma$ discovery of the $pp\rightarrow pWWp$ process could be achieved 
with about 5\invfb{} of data in the leptonic mode only. The signal significance
is calculated as the $P$-value $\alpha$, i.e. as the probability to find the number of 
observed events or more from the background alone.
For 5$\invfb$, the confidence $1-\alpha$ expressed in the number of standard 
deviations for the Gaussian distribution reads 5.3, 5.8, 6.2 for 
$\xi_{max}=0.15,\,0.1,\,0.05$, respectively. The number of signal and 
background events for 5$\invfb$ and 10$\invfb$ together with the value of the 
confidence level, is given in \reftab{anom:measureww_signif}.

It should be noted that the process $pp\rightarrow pWWp$ can be discovered 
even with lower luminosity if one takes the full-leptonic and semi-leptonic 
decays of the two final states $W$ into account. In \cite{Kepka:2008yx} we 
considered 
a simplified analysis studying the two-photon $WW$ production and the 
DPE$\rightarrow WW$ background only assuming  that the overlaid background due 
to multiple interactions is removed with timing detectors. Events with at 
least one lepton above $p^{lep1}_T>30\GeV$ in addition to both proton tags in 
forward detectors $0.0015<\xi_{1,2}<0.15$ were selected. The full-hadronic 
$W$ decays were rejected in order to remove the high QCD dijet background. It 
turned out that the process can be discovered already with $400\invpb$ of 
integrated luminosity by observing 11 signal events and 0.9 background 
yielding a confidence 5.8. The higher sensitivity to the 
two-photon 
$WW$ production is of course due to the higher cross section when one takes 
into account the semi-leptonic decays.
In this case, however, a new background arises from the central exclusive 
production of two quarks which was not studied. If one of the quarks radiates 
a $W$ boson, the $W$+jet+jet final state mimics the semi-leptonic $WW$ decays 
in two-photon production. This background process is planned to be included 
in future releases of FPMC to allow a complete study of the two-photon $WW$ 
production even in the semi-leptonic decay mode~\cite{khozeJames}.

\begin{table}
\centering
\begin{tabular}{|c|cc|ccc|}
\hline
$\xi_{max}$ & signal [fb]     & background [fb] & $S/\sqrt{B+1}$  & \lumi=5\invfb   & \lumi=10\invfb    \\
\hline
\hline
0.05  &      1.69   &   0.06    & & 7.5  & 14  \\
0.1   &      1.57   &   0.05    & &  7.1 & 13    \\
0.15  &      1.07   &   0.02     & &   5.1  & 9.1  \\
\hline
\end{tabular}
\caption{Signal and total background cross sections for $\gamma\gamma\rightarrow  WW$, and the $S/\sqrt{B+1}$ ratio
for luminosities 5 and 10\invfb{} as a function of the forward detector acceptance 0.0015$<\xi<\xi_{max}$ after all
cuts mentioned in the text.}
\label{tab:anom:measureww_signif}
\end{table}

%%%%%%%%%%%%%%%%%%%%%%%%%%%%%%%%%%%%%%%%%%%%%%%%%%%%%%%%%%%%%%%%%%
\section{Sensitivity to quartic anomalous coupling of $W$ and $Z$ to photon}
%%%%%%%%%%%%%%%%%%%%%%%%%%%%%%%%%%%%%%%%%%%%%%%%%%%%%%%%%%%%%%%%%%

%%%%%%%%%%%%%%%%%%%%%%%%%%%%%%%%%%%%%%%%%%%%%%%%%%%%%%%%%%%%%%%%%%
\subsection{Signal cross section for quartic couplings}
%%%%%%%%%%%%%%%%%%%%%%%%%%%%%%%%%%%%%%%%%%%%%%%%%%%%%%%%%%%%%%%%%%
In this section, we study the phenomenological consequences of the new anomalous
terms 
in the Lagrangian. The implementation in the FPMC 
generator allowed us to compare the 
studied signal due to anomalous couplings directly 
with all backgrounds that leave the proton intact and create two leptons, 
electrons or muons, in the central detector.

As shown in \reffig{anom:totalxsectionQGC}, we recall that the anomalous couplings in 
$pp\rightarrow pWWp$ and $pp\rightarrow pZZp$ processes augment the cross 
section from their SM values 95.6\fb{} and 0. The suppression of the cross section due to the 
form factors is shown in \reffig{anom:w2form}. It is important to stress that 
this effect is large and it has to be taken into account when deriving the 
sensitivities to the anomalous couplings.

%%%%%%%%%%%%%%%%%%%%%%%%%%%%%%%%%%%%%%%%%%%%%%%%%%%%%%%%%%%%%%%%%%
\subsection{Background rejection at high luminosity for $WW$ signal}
%%%%%%%%%%%%%%%%%%%%%%%%%%%%%%%%%%%%%%%%%%%%%%%%%%%%%%%%%%%%%%%%%%
\label{kap:highlumi_background}

\begin{figure}
\centering
\includegraphics[width=\twopicwidth]{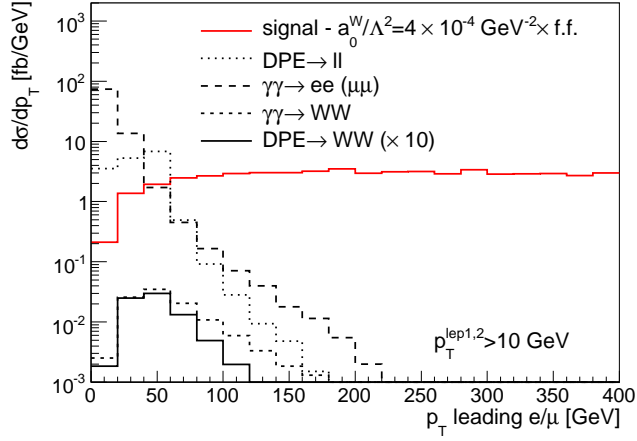}
\caption{Contributions of various background processes to the signal with 
anomalous coupling  $\aZerowLambda=3\times10^{-4}\GeV^{-2}$ with the coupling 
form factors taken into account at generator level.
The signal due to the anomalous coupling manifest itself at high transverse 
lepton momenta.}
\label{fig:anom:ptlead2}
\end{figure}

In \reffig{anom:ptlead2}, the \pt{} distributions of the signal due to quartic 
couplings and the background are superimposed. As expected, the signal due to 
anomalous coupling appears at high transverse momentum, or at high masses.  
The first cut used in the analysis is
therefore to select high $\pt$ leptons together 
with intact protons in the final state detected in the forward detectors to identify
the exclusive two-photon events.
At high luminosity, the forward detector acceptance (high cut on $\xi<0.15$) 
removes the highest mass events and part of the 
signal due to anomalous coupling which appears at high masses is not observed. 

\begin{figure}
\includegraphics[width=\twopicwidth]{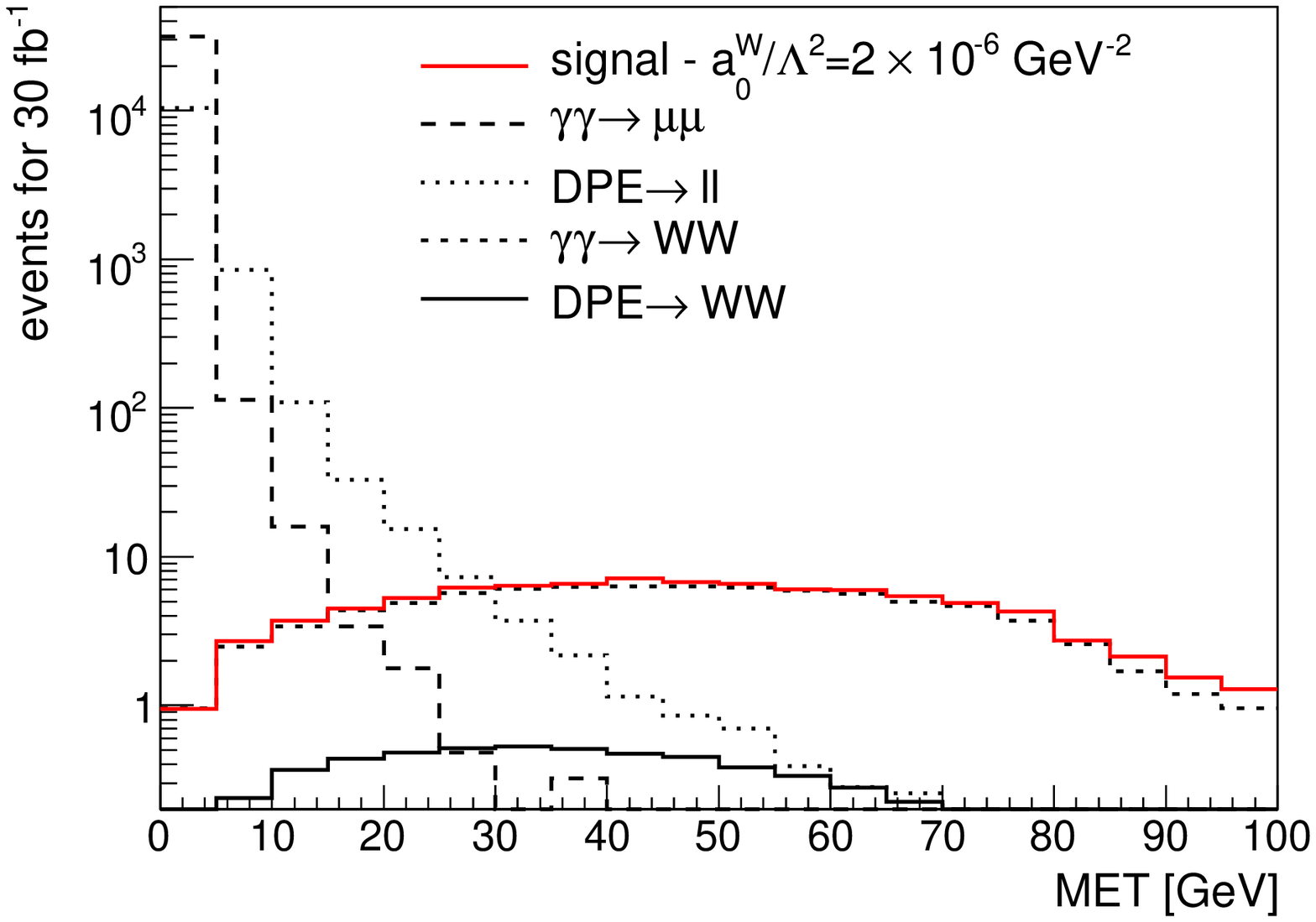}
\includegraphics[width=\twopicwidth]{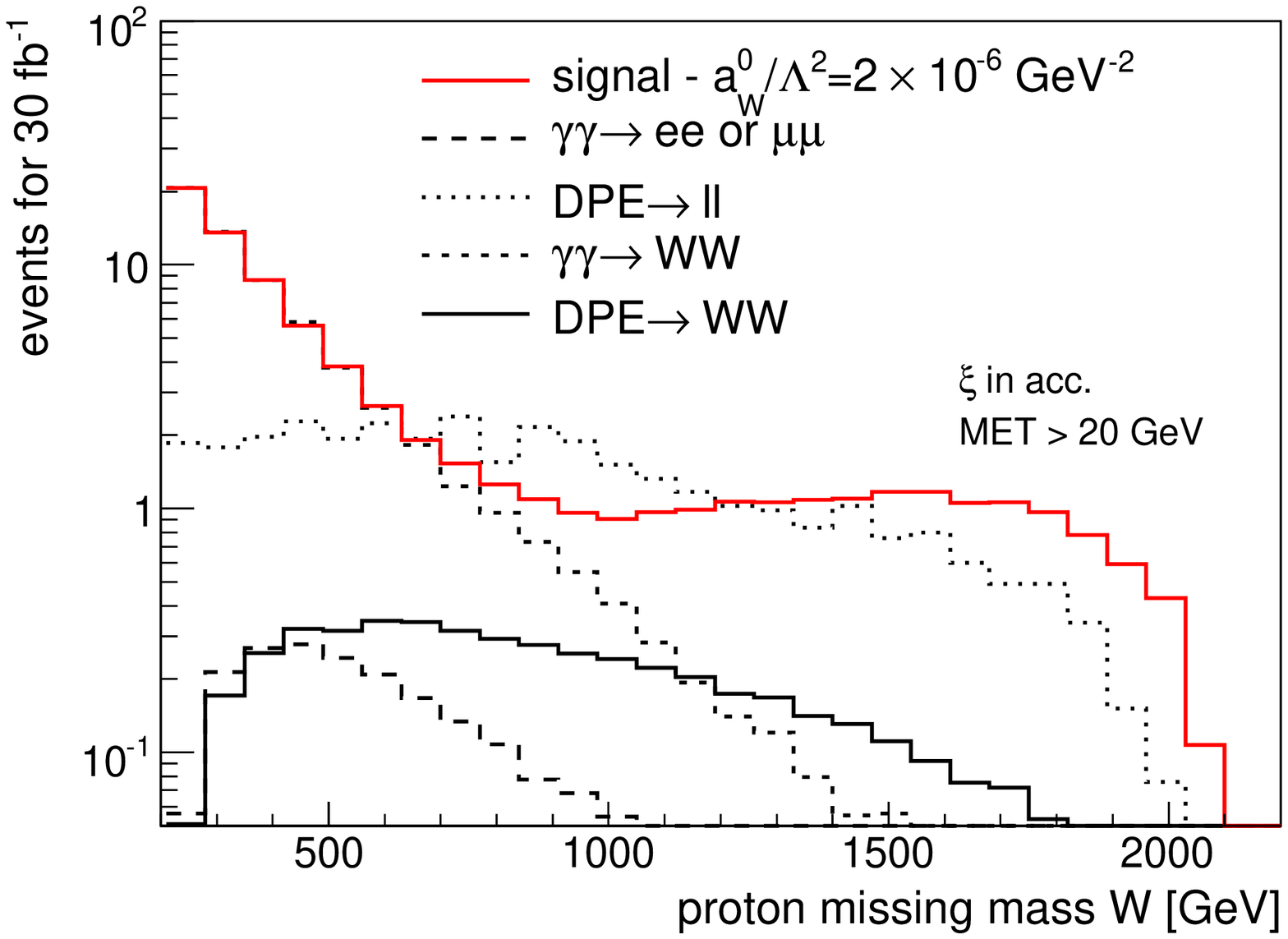}
\caption{Missing transverse energy distribution $\missET$ in the AFP detector 
acceptance cut (left) and proton missing mass (right) in the AFP acceptance 
and  after the cut on $\missET>20\GeV$ cut for signal and all backgrounds 
with \lumi=30\invfb.}
\label{fig:anom:met_w}
\end{figure}
\begin{figure}
%\parbox{\textwidth}{

%\centering
\begin{tabular}{m{\twopicwidth}m{\twopicwidth}}
\includegraphics[width=\twopicwidth]{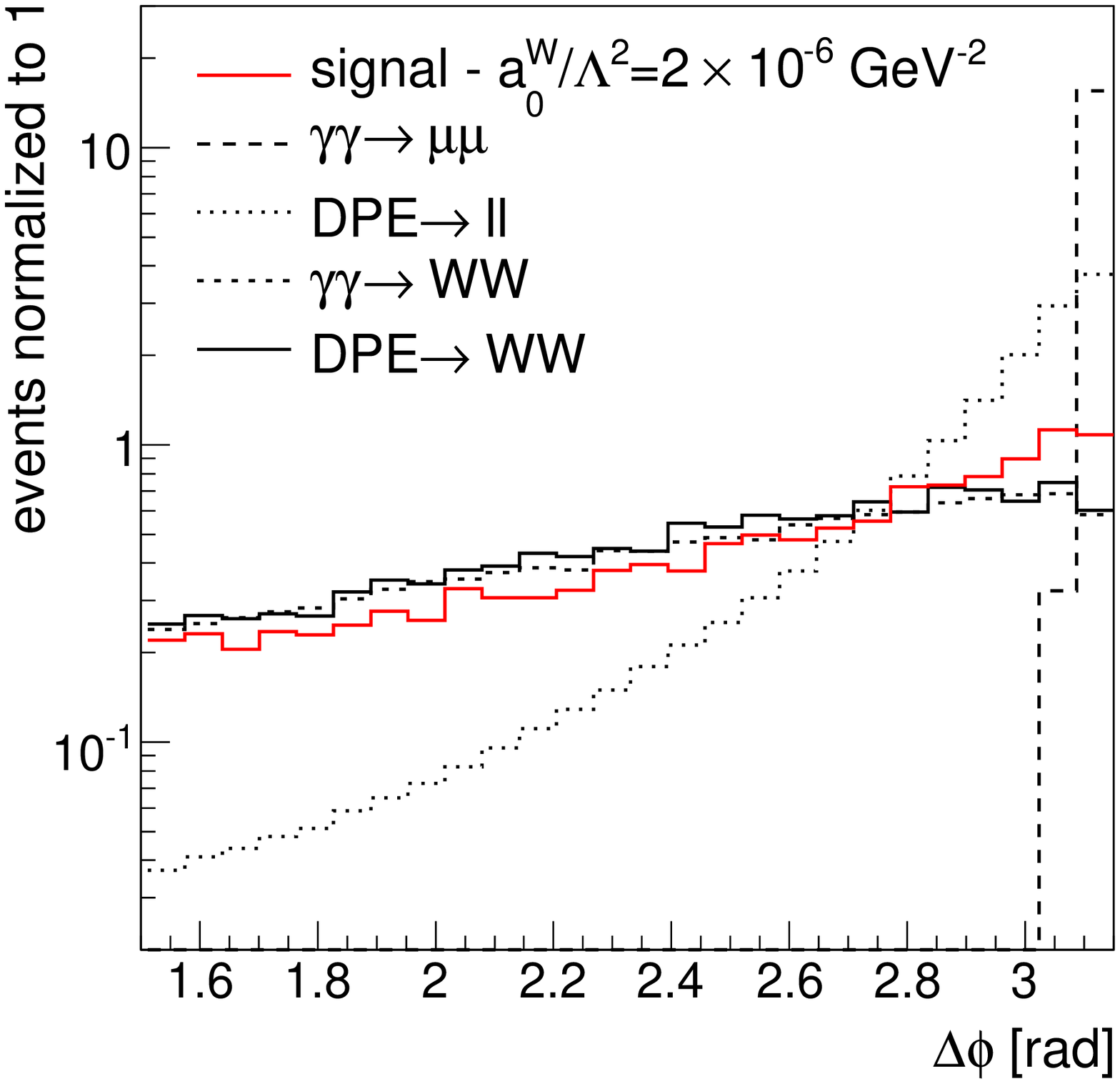} &
\includegraphics[width=\twopicwidth]{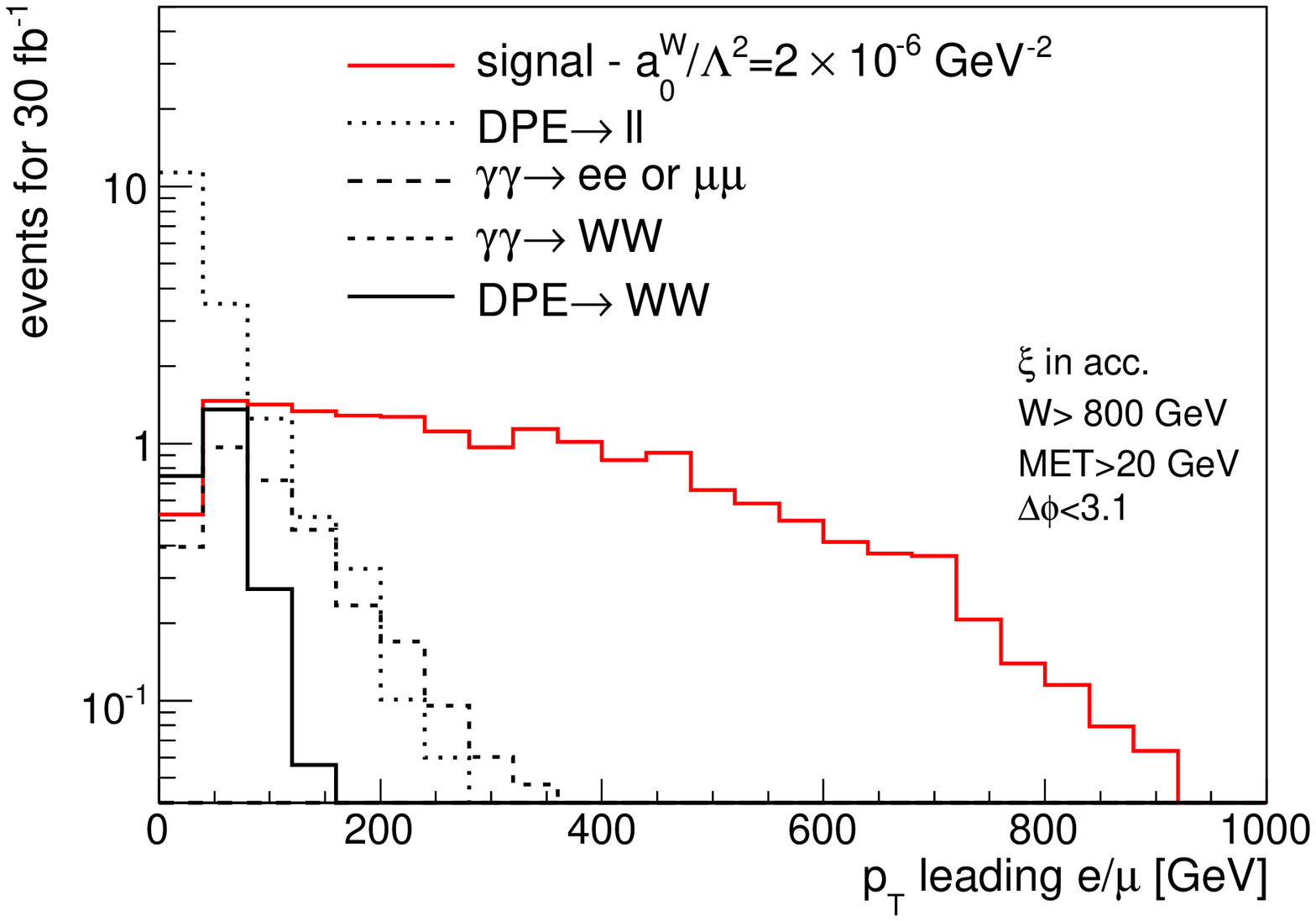}
%}
\end{tabular}
\caption{Angle between detected leptons (left) and $\pt$ distribution  of the 
leading lepton (right) after all cuts as mentioned in the text  for signal 
and background. The events are normalized for \lumi=30\invfb.}
\label{fig:anom:dphi_hl}
\end{figure}

\begin{table}
\begin{tabular}{|c|cccc|cc|c}
\multicolumn{7}{c}{events for 30\invfb} \\
\hline
cut / process &  $\gamma\gamma\rightarrow ee$ & 
$\gamma\gamma\rightarrow \mu\mu$ & $\gamma\gamma\rightarrow \tau\tau$ &
$\gamma\gamma\rightarrow WW$ & DPE$\rightarrow ll$ & DPE$\rightarrow WW$\\
\hline
\hline
gen. $p^{lep1}_T>5\GeV$ & 364500 & 364500   &  337500  & 1198 & 295200  & 530\\
$p^{lep1,2}_T>10\GeV$ &   24895 & 25547 & 177 & 99 & 18464 &  8.8   \\
$0.0015<\xi<0.15$ &  10398   & 10534 & 126 & 89 & 11712 &  6.0 \\
$\missET>20\GeV$    &      0  &     0.86&          14&        77&         36&        4.7   \\
 $W>800\GeV{}$  &    0  &       0.27    &  0.15  &      3.2 &       16    &    2.5  \\
$M_{ll}\notin <80,100>$   & 0   &      0.27 &  0.15      &     3.2 &       13    &  2.5 \\
$\Delta\phi<3.13\rad$   &    0  &         0    &        0.10   &   3.2    &  12  &    2.5   \\
$p_T^{lep1} >160\GeV$ &  0       &       0           &   0      & 0.69    &   0.20  &    0.024   \\
\hline
\end{tabular}
\caption{Rejection of the background by the successive application of the 
selection cuts. The number of events is normalized to $\lumi=30\invfb$ of 
integrated luminosity. The lepton index $lep$ corresponds to electrons or 
muons. The DPE$\rightarrow ll$ was generated with a minimum Drell-Yan mass 
10\GeV. The largest statistical uncertainty is 7\% for DPE$\rightarrow ll$ 
after all cuts.}
\label{tab:anom:nevents_cuts_hl}
\end{table}

\begin{table}
\centering
\begin{tabular}{|c|cc|}
\multicolumn{3}{c}{events for 30\invfb} \\
\hline
cut / couplings (with f.f.)  & $\left|\aZerowLambda\right|=5.4\cdot10^{-6}\,
\GeV^{-2} $   &    $\left|\aCwLambda\right|=20\cdot10^{-6}\,\GeV^{-2}$     \\
\hline
\hline
%gen. $p^{lead. e/\mu}_T>5\GeV$ & 364500 & 364500   &  337500  & 1198 \\
$p^{lep1,2}_T>10\GeV$ &     202   &   200     \\
$0.0015<\xi<0.15$ & 	    116 &    119      \\
$\missET>20\GeV$    &       104  &    107     \\
 $W>800\GeV{}$  &           24  &   23        \\
$M_{ll}\notin <80,100>$ &   24  &   23          \\
$\Delta\phi<3.13\rad$   &        24 &    22           \\
$p^{lep1}_T >160\GeV$ & 	 17  &    16  \\  
\hline
\end{tabular}
\caption{Selection of the signal by the successive application of the cuts. 
The number of events is given for integrated luminosity of $\lumi=30\invfb$. 
The lepton index $lep$ corresponds to electrons or muons.}
\label{tab:anom:nevents_cuts_hl_signal}
\end{table}

The $WW$ events which give a hit in both forward detectors are first selected 
with $\missET>20\GeV$. The $\missET$ dependence
is depicted in \reffig{anom:met_w} (left) for the signal 
$\aZerowLambda=2\times10^{-6}\GeV^{-2}$ and the background. Note that 
the signal is barely distinguishable from the SM $\gamma\gamma\rightarrow WW$ 
process. On the other hand, processes in 
which lepton pairs are created directly through $\gamma\gamma$ or DPE exchange 
are greatly suppressed. The next cut focuses 
on the high diphoton mass $W_{\gamma\gamma}$ where the signal is preferably 
enhanced. In \reffig{anom:met_w} (right) we see
that the signal due to anomalous coupling is well selected if the 
reconstructed missing mass in the forward detectors is $W>800\GeV$.  It was 
verified that such selection applies
for all anomalous parameters in question in a very similar way, i.e. that the 
$W>800\GeV$ retains the interesting signal 
for a wide range of anomalous parameters. 
\par
The most dominant background which remains is the DPE$\rightarrow ll$ production. 
A large part of this background is removed by requesting the angle between 
reconstructed leptons $\Delta\phi<3.13\rad$ as illustrated 
in \reffig{anom:dphi_hl} (left). This removes also the potential two-photon 
dileptons. However, the $\Delta\phi$
cut cannot be arbitrarily relaxed because we would remove part of the signal 
also. We also require the dilepton mass to be far from the $Z$ pole in order to reduce the 
DPE$\rightarrow ll$ production.
\par
To summarize, the following cuts are 
used to select the anomalous signal at high luminosity
  \begin{eqnarray}
 &&p_T^{lep1} >160\GeV,\,p^{lep2}_T>10\GeV,\, 
           0.0015<\xi<0.15,\,
           \missET>20\GeV,\,
            W>800\GeV{},\nonumber\\ 
        &&  M_{ll}\notin \left<80,100\right> \GeV,\,   
           \Delta\phi<3.13\rad          
\end{eqnarray}
Finally, the $p_T^{lep}$ distribution after all mentioned constrains 
is shown in \reffig{anom:dphi_hl} (right). The remaining background is 
composed not only from the expected $\gamma\gamma\rightarrow WW$ production 
but also from DPE$\rightarrow ll$ by about an equal amount.

The successive effect of all cuts and their rejection power of the background 
is summarized in  \reftab{anom:nevents_cuts_hl} where the number of events is 
shown for $\lumi=30\invfb$. The total number of 
background events is thus reduced to $0.90\pm0.05$.

%%%%%%%%%%%%%%%%%%%%%%%%%%%%%%%%%%%%%%%%%%%%%%%%%%%%%%%%%%%%%%%%%%
\subsection{Background rejection at high luminosity for the $ZZ$ signal}
%%%%%%%%%%%%%%%%%%%%%%%%%%%%%%%%%%%%%%%%%%%%%%%%%%%%%%%%%%%%%%%%%%
The $Z$-pair production is background free in the leading order provided that the non-diffractive
background is removed using the forward detectors tagging the intact protons. The complete set of used cuts is 
\be
[(n_{lep}\geq2,\ 2\,\mathrm{of \ same\ charge})\ \mathrm{or}\ n_{lep}\geq3,],\,
0.0015<\xi<0.15,\,
p_T^{lep1}>160\GeV,\,
p_T^{lep2}>25\GeV
\ee

%%%%%%%%%%%%%%%%%%%%%%%%%%%%%%%%%%%%%%%%%%%%%%%%%%%%%%%%%%%%%%%%%%
\subsection{Sensitivity at high luminosity}

\begin{table}
\centering
\begin{tabular}{rl|cccc|}
 & &\multicolumn{4}{c|}{limits $[10^{-6} \GeV^{-2}]$}\\
&form factor             &  $\left|\aZerowLambda\right|$   &    
$\left|\aCwLambda\right|$      &    $\left|\aZerozLambda\right|$  & 
$\left|\aCzLambda\right|$ \\
\cline{2-6}
\multirow{2}{*}{95\% c.l $\Big\{$}& $ \Lambda_{cut}=\infty$  &  1.2  & 4.2  & 2.8   & 10\\
&$ \Lambda_{cut}=2\units{TeV}$      & 2.6    & 9.4 &  6.4   & 24 \\
\cline{2-6}
\multirow{2}{*}{$3\sigma$ evidence $\Big\{$}&$ \Lambda_{cut}=\infty$&  1.6  &  5.8  & 4.0   & 14  \\
&$ \Lambda_{cut}=2\units{TeV}$      & 3.6   & 13  & 9.0   & 34\\ 
\cline{2-6}
\multirow{2}{*}{$5\sigma$ discovery $\Big\{$}&$ \Lambda_{cut}=\infty$ & 2.3  & 9.7  &  6.2 &  23      \\
&$ \Lambda_{cut}=2\units{TeV}$        & 5.4   &  20  & 14  &  52 \\ 
\end{tabular}
\caption{95\% CL interval, 3$\sigma$ evidence, and $5\sigma$ discovery potential
on the $\wwgammagamma$ and $\zzgammagamma$ anomalous quartic parameters 
using \lumi=$30\invfb$ of data at high luminosity with forward detectors, and with or without the form factors applied. 
}
\label{tab:anom:limits_hl30}
\end{table}

\begin{table}
\centering
\begin{tabular}{rl|cccc|}
 & &\multicolumn{4}{c|}{limits $[10^{-6} \GeV^{-2}]$}\\
&form factor             &  $\left|\aZerowLambda\right|$   &    
$\left|\aCwLambda\right|$      &    $\left|\aZerozLambda\right|$  & 
$\left|\aCzLambda\right|$ \\
\cline{2-6}
\multirow{2}{*}{95\% c.l $\Big\{$}& $ \Lambda_{cut}=\infty$  &  0.7  & 2.4  &  1.1  & 4.1\\
&$ \Lambda_{cut}=2\units{TeV}$      &  1.4   & 5.2 & 2.5    & 9.2 \\
\cline{2-6}
\multirow{2}{*}{$3\sigma$ evidence $\Big\{$}&$ \Lambda_{cut}=\infty$&  0.85  & 3.0   & 1.6   & 5.7  \\
&$ \Lambda=2\units{TeV}$      &  1.8  & 6.7  &  3.5  &  13\\ 
\cline{2-6}
\multirow{2}{*}{$5\sigma$ discovery $\Big\{$}&$ \Lambda_{cut}=\infty$ &  1.2 &  4.3 & 4.1  &  8.9      \\
&$ \Lambda_{cut}=2\units{TeV}$        & 2.7   &   9.6 & 5.5  & 20 \\ 
\end{tabular}
\caption{95\% CL interval, 3$\sigma$ evidence, and $5\sigma$ discovery potential
on the $\wwgamma\gamma$ and $\zzgammagamma$ anomalous quartic parameters 
using \lumi=$200\invfb$ of data at high luminosity with forward detectors, and with or without the form factors applied. 
95\% CL limit, 3$\sigma$ evidence, and $5\sigma$ discovery potential correspond to the values of the couplings outside of the quoted intervals. 
}
\label{tab:anom:limits_hl200}
\end{table}

\begin{figure}
\includegraphics[width=\twopicwidth]{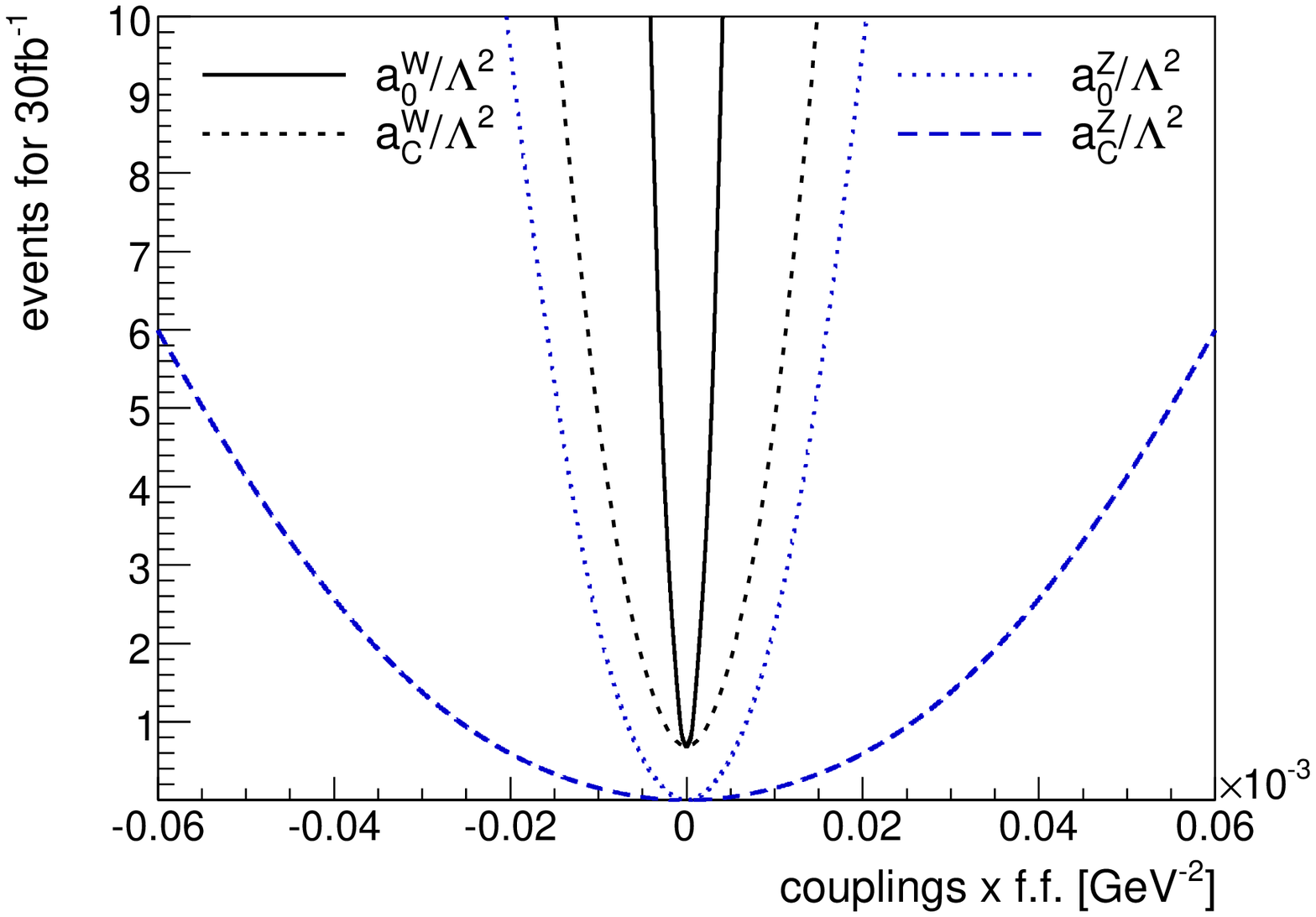}
\includegraphics[width=\twopicwidth]{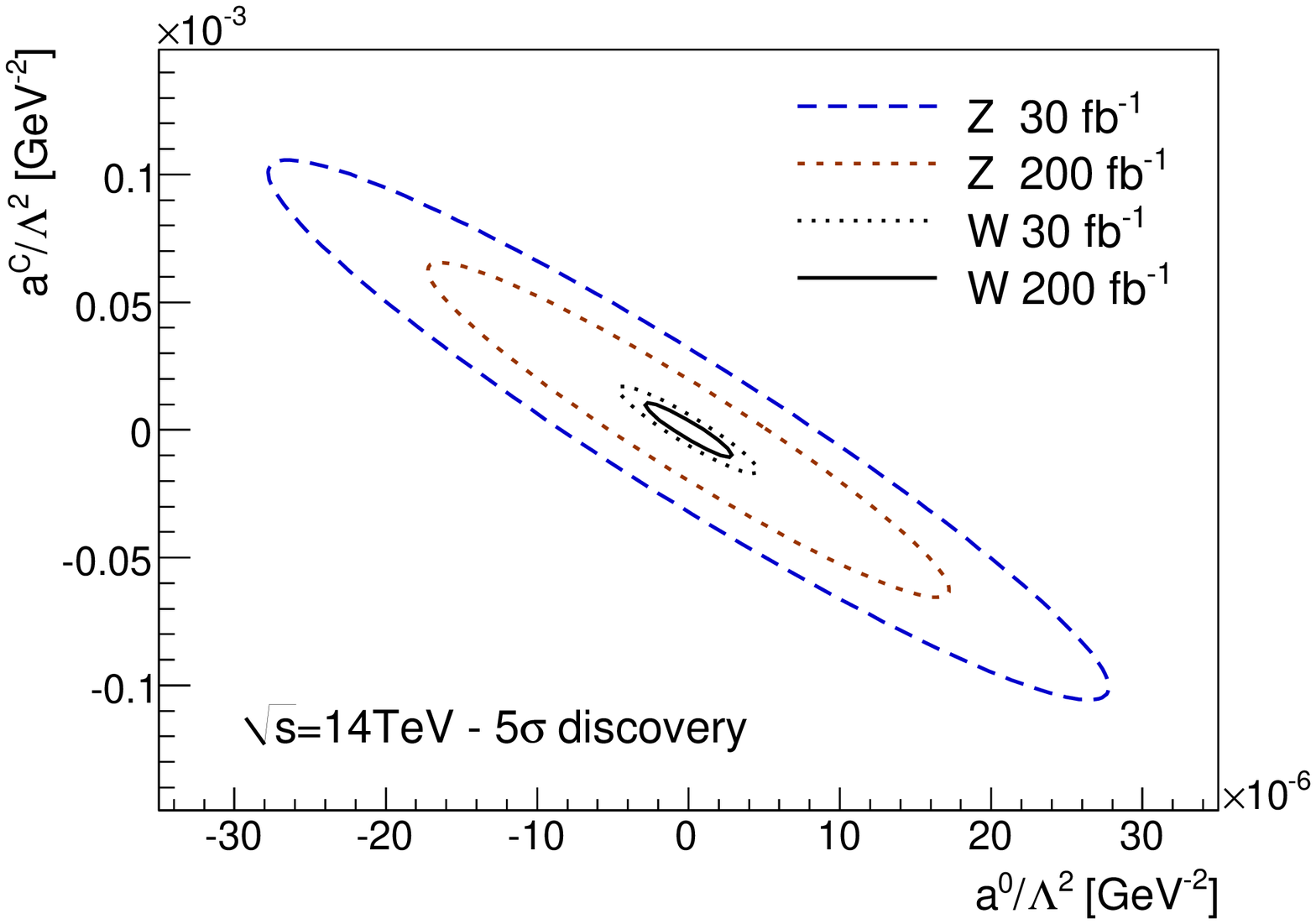}
\caption{Number of events for signal (left) due to different values of 
anomalous couplings after all cuts (see text) for \lumi=30\invfb, and 
$5\sigma$ discovery contours (right) for all the $WW$ and $ZZ$ quartic 
couplings at $\sqrt{s}=14\TeV$ for \lumi=30\invfb{}and \lumi=200\invfb.}
\label{fig:anom:limits_hl}
\end{figure}
%%%%%%%%%%%%%%%%%%%%%%%%%%%%%%%%%%%%%%%%%%%%%%%%%%%%%%%%%%%%%%%%%%

The number of events after all cuts as a function of the anomalous parameters shown in Figure \reffig{anom:limits_hl} (left) is used
to calculate the exclusion upper limits. The results are summarized
in Tables \ref{tab:anom:limits_hl30} and \ref{tab:anom:limits_hl200} for 
$\lumi=30\invfb$ and $\lumi=200\invfb$, respectively.
\par

Comparing our results with the OPAL limits \refeq{anom:qgclimits} we 
see that the improvement of sensitivities which can be obtained 
with a collected luminosity $30\invfb$ corresponding approximately to three 
years of running with the forward detectors, is about a factor of 5000 for 
all couplings except $\aCzLambda$
where the improvement is about a factor 5 worse. With the 
full \lumi=200\invfb{} luminosity, the improvement reaches about a factor of 
10000. When two of the anomalous parameters are varied independently, 
the sensitivities form ellips $a_{C}/\Lambda^2\times a_{0}/\Lambda^2$ plane shown in 
\reffig{anom:limits_hl} (right).

%%%%%%%%%%%%%%%%%%%%%%%%%%%%%%%%%%%%%%%%%%%%%%%%%%%%%%%%%%%%%%%%%%
\section{Sensitivity to anomalous triple gauge $\wwgamma$ coupling
at high luminosity}
%%%%%%%%%%%%%%%%%%%%%%%%%%%%%%%%%%%%%%%%%%%%%%%%%%%%%%%%%%%%%%%%%%

%%%%%%%%%%%%%%%%%%%%%%%%%%%%%%%%%%%%%%%%%%%%%%%%%%%%%%%%%%%%%%%%%%
\subsection{Signal selection}
%%%%%%%%%%%%%%%%%%%%%%%%%%%%%%%%%%%%%%%%%%%%%%%%%%%%%%%%%%%%%%%%%%
The limits on triple gauge boson anomalous couplings obtained at LEP and the Tevatron are already very stringent, more 
than in the case of quartic anomalous couplings.
Let us however remind that triple and genuine quartic anomalous couplings are not 
related in any way. Hence, the analysis which has 
been performed above for the quartic couplings has its own importance 
irrespective of the triple ones.
The production cross sections corresponding to the current limits for $\dkap$ 
and $\lam$ are rather small, hence, the only option to gain an 
improvement is to consider the high luminosity scenario with forward detectors.

The signal selection follows closely two already defined strategies. Since 
$\dkap$ changes only the normalization, 
the signal at low $W$  masses has to be retained. Therefore the 
selection of the signal is the same as it was optimized for the  measurement 
of the SM $pp\rightarrow p WW\!p$ cross section (\refsec{kap:anom:measurement}). On the contrary, the signal due to $\lam$ parameters appears at high mass with high $\pt$ objects created in the central detector. We can simply use the signal selection requirements designed for the quartic couplings discussed in  (\refsec{kap:highlumi_background}). For clarity, we use the following cuts:\\ 
for $\dkap$:
  \begin{eqnarray}
p_T^{lep1}>25\GeV,\,
 p^{lep2}_T>10\GeV,\,	      
 0.0015<\xi<0.15,\,	     
 \missET>20\GeV,\,    
 160<W<500\GeV,\,
 \Delta\phi<2.7\rad
\end{eqnarray}
and for $\lam$:
  \begin{eqnarray}
 &&p_T^{lep1} >160\GeV,\,p^{lep2}_T>10\GeV,\, 
           0.0015<\xi<0.15,\,
           \missET>20\GeV,\,
            W>800\GeV{},\nonumber\\ 
        &&  M_{ll}\notin \left<80,100\right> \GeV,\,   
           \Delta\phi<3.13\rad         
\end{eqnarray}

The expected backgrounds for $\lumi=30\invfb$ are $1.7\pm0.1$\fb{} for $\dkap$ 
and $0.90\pm0.05$ for $\dkap$ as discussed in 
sections \ref{kap:anom:measurement} and \ref{kap:highlumi_background}. The 
successive application of all mentioned requirements for $\dkap$ and $\lam$ 
signal is detailed in \reftab{anom:nevents_cuts_hl_signaltriple} for \lumi=30\invfb.

 \begin{table}
\centering
\begin{tabular}{|c|c|}
\multicolumn{2}{c}{events for 30\invfb} \\
\hline
 cut   & $ \dkap=0.3$ (with f.f.)         \\
\hline
\hline
  $p^{lep1,2}_T>10\GeV$  &	    194 \\      
 $0.0015<\xi<0.15$ &	          179  \\
 $\missET>20\GeV$ &            158    \\
 $W>160\GeV$ &                 158   \\
 $\Delta\phi<2.7\rad$ &            118  \\
 $p_T^{lep1}>25\GeV$ &         112 \\
 $W<500$   &                   98 \\
\hline
\end{tabular}
\begin{tabular}{|c|c|}
\multicolumn{2}{c}{events for 30\invfb} \\
\hline
cut   & $ \lam=0.1$ (with f.f.)         \\
\hline
\hline
%gen. $p^{lead. e/\mu}_T>5\GeV$ & 364500 & 364500   &  337500  & 1198 \\
$p^{lep1,2}_T>10\GeV$ 	&              168.              \\
$0.0015<\xi<0.15$	 & 	         119            \\
$\missET>20\GeV$    	&                107                \\
 $W>800\GeV{}$  	&                25               \\
$M_{ll}\notin <80,100>$ &             25                        \\
$\Delta\phi<3.13\rad$   	&                24                      \\
$p_T^{lep1} >160\GeV$ &        19                            \\
\hline
\end{tabular}
\caption{Selection of the $\dkap$ and $\lam$ signal by the successive 
application of the cuts. The number of events is given for integrated 
luminosity $\lumi=30\invfb$.}
\label{tab:anom:nevents_cuts_hl_signaltriple}
\end{table}

\begin{table}
\centering
\begin{tabular}{r|cc|cc}
\multicolumn{1}{c}{}&\multicolumn{2}{c}{$\lumi=30\invfb$}&\multicolumn{2}{c}{$\lumi=200\invfb$}\\
\multicolumn{1}{c}{}  &$\dkap$  &   $\lam$  &    $\dkap$  &   $\lam$        \\
\cline{2-5}
 \multirow{1}{*}{95\% c.l $\big\{$}   &  [-0.25, 0.16]     &  [-0.052, 0.049]  &    [-0.096, 0.057] &    [-0.023, -0.027]   \\
\cline{2-5} 
\multirow{1}{*}{$3\sigma$ evidence $\big\{$}   &   [-0.39, 0.25]  &  [-0.066, 0.064]  &    [-0.136, 0.087]&    [-0.037, 0.038] \\ 
\cline{2-5}
\multirow{1}{*}{$5\sigma$ evidence $\big\{$}      &   [-0.67, 0.40]  &  [-0.088, 0.094]  &    [-0.26, 0.16] &    [-0.053, 0.049]\\ 
\end{tabular}                                            
\caption{95\% CL, 3$\sigma$ evidence, and $5\sigma$ discovery potential on 
the $\wwgamma$ 
anomalous parameters for a luminosity of \lumi=30$\invfb$ and 200$\invfb$ 
using the AFP forward detectors with coupling form factors applied.}
\label{tab:anom:limits200}
\end{table}

%%%%%%%%%%%%%%%%%%%%%%%%%%%%%%%%%%%%%%%%%%%%%%%%%%%%%%%%%%%%%%%%%%
\subsection{Sensitivities at high luminosities}
%%%%%%%%%%%%%%%%%%%%%%%%%%%%%%%%%%%%%%%%%%%%%%%%%%%%%%%%%%%%%%%%%%

The sensitivities are 
summarized in \reftab{anom:limits200} for 30 and 200\invfb. Comparing these 
values  with the current limits from the Tevatron, we see that the improvement 
is limited, about a factor of 2 with 30\invfb of collected luminosity.

Let us also compare the results to those obtainable in the conventional ATLAS 
analysis without forward detectors. $WW\gamma$ anomalous couplings are probed 
by fitting the $p_T^\gamma$ spectrum of the photon distribution to the NLO 
expectation using the combined sample of $W(e\nu)\gamma$ and $W(\mu\nu)\gamma$ 
events or by fitting the transverse mass distribution $M_T(WW)$ of the boson 
pair, reconstructed from the two observed leptons and the missing transverse 
energy \cite{Aad:2009wy}. The corresponding 95\% CL limits obtained for 
$\lumi=30\invfb$, with the same form factor assumption as 
before \refeq{explicit_formfactor} are shown in \reftab{anom:tgc_atlas}. The 
presented analysis using forward detectors has about a factor 2 worse  
precision than the analysis in non-diffractive studies and would therefore be 
a complementary measurement.

The disadvantage of the full leptonic ($e/\mu$) channel of the boson decays is 
the small rates since only $\thickapprox4\%$ of the signal 
is kept. In the work presented in \cite{Kepka:2008yx}, we performed a quite 
simple analysis (without simulating all possible backgrounds)
assuming that  $\gamma\gamma\rightarrow WW$ and 
DPE$\rightarrow WW$ are the only important backgrounds, but keeping also the 
semi-leptonic events. 
The improvement for $\lam$ with respect to the analysis with leptonic decays 
is only modest, since the selection was not optimized for high masses where 
the signal appears. On the other hand, the larger signal sample when 
semi-leptonic decays are included yields a better separation of the signal due 
to the $\dkap$ anomalous parameter with respect to the SM 
$\gamma\gamma\rightarrow WW$ production and the sensitivity is improved by a 
factor 4. However, the full study using the semi-leptonic decays 
and all simulated backgrounds will be
performed in an incoming paper. Especially, we still need
to implement one additional background in FPMC
due to the central exclusive production of $q\bar{q}$ pairs  in which one
of the quarks radiates a $W$ boson. This process has not been considered in the 
phenomenological studies of this kind so far.

\section{Conclusion}
In this paper, it was first shown  how the SM two-photon production 
$pp\rightarrow pWW\!p$ process with both $W$s 
decaying leptonically could be selected  from the diffractive or exclusive 
background. Using the forward detectors, about 50 events can be  observed with 
30\invfb{} of collected luminosity corresponding approximately to 3~years of 
data taking whereas the number of background processes could
be kept at a few events level. No multiple interaction background was studied, 
but the  boson invariant mass $2\times m_W$ threshold could be used to 
suppress this background using the AFP proton tagging (in addition, the proton arrival time measured with special fast timing detectors can be used 
to further suppress the overlap background). 
\par 
The sensitivities to triple and quartic gauge anomalous couplings 
in $W$ production via photon induced processes were studied 
using the standalone ATLFast++ simulation. To reduce the number of
background events for this first study,
only leptonic decays of the $W$s were considered, and the case of
the semi-leptonic decays will be the subject of an incoming paper.
Using a high luminosity of 
30 or $200\invfb$ with 
the forward detectors to tag the exclusive two-photon events, the sensitivities
to the quartic couplings can be improved by more than four orders of magnitude. 
\par
On the other hand, the improvement of the triple gauge coupling experimental 
constraints is smaller. In the full-leptonic channel, the $\dkap$  analysis 
cannot yield better results than the current limits coming from OPAL; however, it can give 
better results than those from the Tevatron. On the other hand, the $\lam$ 
parameter can be fully constrained by a factor 2 better with respect to the 
OPAL collaboration at LEP.

\begin{table}
\centering
\begin{tabular}{|c|cc|}
\hline
                         &     $\dkap$        &  $\lam$         \\\hline
$W\gamma, (p_T^{\gamma})$&     [-0.11, 0.05]  &  [-0.02, 0.01]  \\
$WW, (M_T)$              &     [-0.056, 0.054]&  [-0.052, 0.100]  \\
\hline
\end{tabular}
\caption{95\% CL limits on the $\wwgamma$\ coupling parameters obtained from 
fitting the
$p_T^{\gamma}$ and $M_T(WW)$ distributions in $W\gamma$ and  $WW$ final states 
in inelastic production
in ATLAS, and calculated for \lumi=$30\invfb$ and for the form factors 
$\Lambda=2$\TeV, $n=2$
\cite{Aad:2009wy}. }
\label{tab:anom:tgc_atlas}
\end{table}


\begin{thebibliography}{99}

\bibitem{stirling} P.~J.~Dervan, A.~Signer, W.~J.~Stirling, A.~Werthenbach,
J. Phys. {\bf G26} (2000);
W.~J.~Stirling, A.~Werthenbach, Eur. Phys. J. {\bf C14} (2000) 103.

\bibitem{higgsless} O. J. P. Eboli, M. C. Gonzales-Garcia,
S. M. Lietti, S. F. Novaes, Phys. Rev. {\bf D63} (2001) 075008;
G. Cvetic, B. Koegerler, Nucl. Phys. {\bf B363} (1991)  no2-3,401-424;
A. Hill, J.J. van der Bij, Phys. Rev. {\bf D36} (1987) 3463;
 

\bibitem{piotr} J. de. Favereau et al., preprint arXiv:0908.2020.

\bibitem{fpmc}
M. Boonekamp, V. Jur\'{a}nek, O. Kepka, C. Royon, Forward Physics Monte
Carlo, Proceedings of the Workshop of the Implications of HERA for LHC
physics;
 %H.~Jung {\it et al.},
 %``Proceedings of the workshop: HERA and the LHC workshop series on the
 %implications of HERA for LHC physics,''
 arXiv:0903.3861 [hep-ph];\\
 %%CITATION = ARXIV:0903.3861;%%
 \url{http://cern.ch/fpmc}.
 
\bibitem{Budnev}
  V.~M.~Budnev, I.~F.~Ginzburg, G.~V.~Meledin and V.~G.~Serbo,
  %``The Two photon particle production mechanism. Physical problems.
  %Applications. Equivalent photon approximation,''
  Phys.\ Rept.\  {\bf 15} (1974) 181.
  %%CITATION = PRPLC,15,181;%%

\bibitem{Boonekamp:2007iu}
  M.~Boonekamp, C.~Royon, J.~Cammin and R.~B.~Peschanski,
  %``Threshold scans in diffractive W pair production via QED processes at the
  %LHC,''
  Phys.\ Lett.\  B {\bf 654} (2007) 104
  [arXiv:0709.2742 [hep-ph]].
  %%CITATION = PHLTA,B654,104;%%
%\cite{Budnev:1974de}


\bibitem{TreelevUnit1} 
  J.~M.~Cornwall, D.~N.~Levin and G.~Tiktopoulos,
  %``Uniqueness of spontaneously broken gauge theories,''
  Phys.\ Rev.\ Lett.\  {\bf 30} (1973) 1268
  [Erratum-ibid.\  {\bf 31} (1973) 572].
  %%CITATION = PRLTA,30,1268;%%

%\cite{Cornwall:1974km}
\bibitem{TreelevUnit2}
  J.~M.~Cornwall, D.~N.~Levin and G.~Tiktopoulos,
  %``Derivation Of Gauge Invariance From High-Energy Unitarity Bounds On The S
  %Matrix,''
  Phys.\ Rev.\  D {\bf 10} (1974) 1145
  [Erratum-ibid.\  D {\bf 11} (1975) 972].
  %%CITATION = PHRVA,D10,1145;%%

%\cite{Lust:2008qc}
\bibitem{electroweakCorrections}
  A.~Denner, S.~Dittmaier and R.~Schuster,
  %``Electroweak radiative corrections to gamma gamma --> W+ W-,''
  arXiv:hep-ph/9601355.
  %%CITATION = HEP-PH/9601355;%%

%\cite{Amsler:2008zzb}
\bibitem{Amsler:2008zzb}
  C.~Amsler {\it et al.}  [Particle Data Group],
  %``Review of particle physics,''
  Phys.\ Lett.\  B {\bf 667}, 1 (2008).
  %%CITATION = PHLTA,B667,1;%%


\bibitem{misha}
  V.~Khoze and M.~Ryskin, private communication.

\bibitem{Belanger:1992qh}
  G.~Belanger and F.~Boudjema,
  %``Probing Quartic Couplings Of Weak Bosons Through Three Vectors Production
  %At A 500-Gev Nlc,''
  Phys.\ Lett.\  B {\bf 288}, 201 (1992).
  %%CITATION = PHLTA,B288,201;%%

%\cite{Abbiendi:2004bf}
\bibitem{LEPlimitsQGC}
  G.~Abbiendi {\it et al.}  [OPAL Collaboration],
  %``Constraints on anomalous quartic gauge boson couplings from nu anti-nu
  %gamma gamma and q anti-q gamma gamma events at LEP2,''
  Phys.\ Rev.\  D {\bf 70} (2004) 032005
  [arXiv:hep-ex/0402021].
  %%CITATION = PHRVA,D70,032005;%%

%\cite{Eboli:2000ad}
\bibitem{Eboli:2000ad}
  O.~J.~P.~Eboli, M.~C.~Gonzalez-Garcia, S.~M.~Lietti and S.~F.~Novaes,
  %``Anomalous quartic gauge boson couplings at hadron colliders,''
  Phys.\ Rev.\  D {\bf 63} (2001) 075008
  [arXiv:hep-ph/0009262].
  %%CITATION = PHRVA,D63,075008;%%


%\cite{Pierzchala:2008xc}
\bibitem{Pierzchala:2008xc}
  T.~Pierzchala and K.~Piotrzkowski,
  %``Sensitivity to anomalous quartic gauge couplings in photon-photon
  %interactions at the LHC,''
  Nucl.\ Phys.\ Proc.\ Suppl.\  {\bf 179-180} (2008) 257
  [arXiv:0807.1121 [hep-ph]].
  %%CITATION = NUPHZ,179-180,257;%%
           
%\cite{Hagiwara:1986vm}
\bibitem{Hagiwara:1986vm}
  K.~Hagiwara, R.~D.~Peccei, D.~Zeppenfeld and K.~Hikasa,
  %``Probing the Weak Boson Sector in e+ e- $\to$ W+ W-,''
  Nucl.\ Phys.\  B {\bf 282}, 253 (1987).
  %%CITATION = NUPHA,B282,253;%%

\
%\cite{Kepka:2008yx}      
\bibitem{Kepka:2008yx}
O.~Kepka and C.~Royon,
%``Anomalous $W W \gamma$ coupling in photon-induced processes using forward
%detectors at the LHC,''
Phys.\ Rev.\  D {\bf 78} (2008) 073005
[arXiv:0808.0322 [hep-ph]].
%%CITATION = PHRVA,D78,073005;%%





%\cite{Alcaraz:2006mx}
\bibitem{LEPlimits}
  \mbox{J.~Alcaraz {\it et al.}}  [LEP Electroweak Working Group],
  %``A combination of preliminary electroweak measurements and constraints on
  %the standard model,''
  arXiv:hep-ex/0612034.
  %%CITATION = HEP-EX/0612034;%%

%\cite{:2008vja}
\bibitem{TEVlimits}
  V.~M.~Abazov {\it et al.}  [D0 Collaboration],
  %``First study of the radiation-amplitude zero in Wgamma production and limits
  %on anomalous WWgamma couplings at sqrt(s)=1.96 TeV,''
  arXiv:hep-ex/0803.0030.
  %%CITATION = ARXIV:0803.0030;%%
%\cite{Denner:1996wm}


\bibitem{herwig} G.~Marchesini {\it et al.}, 
Comp.~Phys.~Comm.~{\bf 67}, 465 (1992).

%\cite{Boos:2004kh}
\bibitem{comphep}
  E.~Boos {\it et al.}  [CompHEP Collaboration],
  %``CompHEP 4.4: Automatic computations from Lagrangians to events,''
  Nucl.\ Instrum.\ Meth.\  A {\bf 534} (2004) 250
  [arXiv:hep-ph/0403113].
  %%CITATION = NUIMA,A534,250;%%

\bibitem{herapdf} 
  A.~Aktas {\it et al.}  [H1 Collaboration],
  Eur.\ Phys.\ J.\  C {\bf 48} (2006) 715;
  A.~Aktas {\it et al.}  [H1 Collaboration],
  Eur.\ Phys.\ J.\  C {\bf 48} (2006) 749;.
  S.~Chekanov  [ZEUS Collaboration],
  Nucl.\ Phys.\  B {\bf 800} (2008) 1.
  

\bibitem{dglap} G.Altarelli and G.Parisi,
{\ Nucl. Phys.} {\bf B126}  18C (1977) 298.
V.N.Gribov and L.N.Lipatov, { Sov. Journ. Nucl. Phys.} (1972) 438 and 675.
Yu.L.Dokshitzer, { Sov. Phys. JETP.} {\bf 46} (1977) 641.

\bibitem{survival} 
  V.~A.~Khoze, A.~D.~Martin and M.~G.~Ryskin,
  Eur.\ Phys.\ J.\  C {\bf 23} (2002) 311;  
  E.~Gotsman, E.~Levin, U.~Maor, E.~Naftali and A.~Prygarin,
  arXiv:hep-ph/0511060.


\bibitem{atlfast} 
   ATLFast++ package for ROOT, \url{http://root.cern.ch/root/Atlfast.html}.
  

\bibitem{afp} M. G. Albrow et al., JINST 4 (2009) T10001; C. Royon,
Proceedings of the DIS 2007 workshop, Munich, preprint arXiv:0706.1796.






            

\bibitem{khozeJames}
   V.~Khoze, W.~J.~Stirling, private communication.


%\cite{Aad:2009wy}
\bibitem{Aad:2009wy}
  G.~Aad {\it et al.}  [The ATLAS Collaboration],
  %``Expected Performance of the ATLAS Experiment - Detector, Trigger and
  %Physics,''
  arXiv:0901.0512 [hep-ex].
  %%CITATION = ARXIV:0901.0512;%%


\end{thebibliography}
\end{document}